\documentclass{aa}

\bibpunct{(}{)}{;}{a}{}{,} 

\usepackage{graphicx}
\graphicspath{ {figures/} }

\usepackage[T1]{fontenc}
\usepackage{textcomp,gensymb}
\usepackage[varg]{txfonts}
\usepackage{natbib}
\usepackage{amsmath}

\title{Zukunftoptik}
\subtitle{reconsidering light and radio astronomy}
\author{E. Barrelet}
\institute{LPNHE, Laboratoire de Physique Nucléaire et de Hautes Énergies, CNRS - IN2P3 - Universités Paris VI et VII,  4, Place Jussieu - Tour 33 - Rez-de-Chaussée 75252 Paris Cedex 05}

\begin{document}

\abstract {Recent progresses of electronics, essentially due to its miniaturization, are opening new fields that were just dreamed 
of, notably in astronomy. At start in §\ref{sec:OPSEC3}, we introduce the time variation of images expressing the dual nature of the
optical signal (ZO) and we expose several useful applications where the optical signal variations are not faster than CCD.
However we prefered to initiate the article with a deeper question posed inadvertently in §\ref{sec:OPSEC2}: what causes the rapid, well timed
and regular variation of the signals induced in our test setup, which we see in Fig. \ref{fig:1} The answer proposed are two causes: one 
is a light photon acting indirectly through the induction of a large number of secondary electrons (§\ref{sec:OPSEC2}), the other are 
the RF photons (subliminal, but acting directly) as detailed in §\ref{sec:OPSEC4}. For both light and RF, using a sum of induced currents instead 
of a single photon quadri-vector transform the case.  

}

\maketitle

\section{Introduction}

This article develops an original vision of astronomy based on a novel Quantum Mechanical (QM) optics and its extension from light to radio.
It introduces concepts that are immediately applicable to experimental data, with a spectacular profit. All this is ultimately made possible 
by the progresses of electronics for measurement and computation in the last fifteen years. Some of these improvements have been long anticipated by
 \citet[p.7]{ref:10}\footnote{Born was both the prominent professor of optics in German universities and the 1952 Nobel for having fathered  
Quantum Mechanics. He also proposed ``Born elastic scattering'' occuring when a photon wavefunction impinges on a condensed matter object, 
such as a mirror or a refractive lens, either concave or convex.}, for which he foresaw the refoundation of optics based on QM, 
under the name of <<Zukunftoptik>> (ZO), alias <<Future optics>>. We learned these historical facts while reading \citet{ref:1} and 
after having presented in \citet{ref:2} a remarquable improvement of the precision of the astronomical measurement of light intensity 
--from $1\%$ to $1ppm$-- by taking into account each photon propagation and its storage in a CCD detector. Actually this took into account
the minute effects due to the photons QM scattering on the microscopic defects of optical surfaces, but not their macroscopic QM elastic 
scattering on large objects such as whole mirrors and their periphery (such as the telescope mirror inner and outer cuts which QM effect 
is seen from a $15m$ distance in Fig.\ref{fig:8}).
Photon wavefunctions issued from a distant point source propagate in vacuum as described by the same Fresnel kernel. 
Two partial fluxes emitted by the same source are linearly correlated and each flux is linearly correlated with its own squared
fluctuations (Fig.\ref{fig:7}).

The optical signal contains two types of ZO components : single photon QM wavefunctions --through an angular beam cross-section-- 
and intensity variation of the emission process --recorded in photon time series along beams--. 
This induces a much larger complexity of the wavefunction after scattering, while the propagation of photons in vacuum remains simple.  
Beyond the quantum angular dependence recorded by CCDs, ZO records the intensity time variations with their statistical fluctuations.
That is done here using the fast digitization of current just achieved a few years ago\footnote{however the nature of the current 
and of the time resolution of its sampling remains an open question}. The sequence of current samples constitutes the intensity time variation 
signal yielding the frequency spectrum seen in Fig. \ref{fig:1}.

In the course of this paper, we shall present two types of light intensity signals, either yielding a stationary spectrum as in Section 
\ref{sec:OPSEC2}, or a time varying spectrogram in Section \ref{sec:OPSEC3}, both limited to 2.5 GHz by the 5 GHz current sampling frequency 
(5.3 orders of magnitude below light's optical frequencies). Such light intensity signals joined to the quantum single photon waveforms and 
to Born elastic scattering constitute the core of ZO resolution for light. Another type of signal generated by ambient Radio-Frequency (RF) 
was found inside the ``noise'' peaks of Fig. \ref{fig:1}. It is produced by various local telecommunication emitters and received by the 
virtual broadband antenna effect in the coaxial line seen in that figure.  
Unexpectedly, this signal presents all the tenuous characteristics of a radio-astronomy signal that we could either study in 
§\ref{subsec:OPSEC4_3} or emulate in §\ref{subsec:OPSEC4_4}. 
In particular it is subliminal, that is below one digitizing unit which is itself below the average current noise. 
This turned out to be an occasion to address the most fundamental questions with an original point of view which unifies the ZO treatment of 
relativistic photons with wavelengths in the range from light to RF (a factor $10^7$). 
The concentration of light and RF photons by paraboloid concave mirrors are treated using Born QM elastic scattering.
The interference and interferometry between duplicated light and RF variable intensity beams are explained by the same statistical correlation 
analysis. Nevertheless, wavelength and  energy per photon are so different in these two domains that their signal detections are not commensurate.  
To conclude this Section \ref{sec:OPSEC4}, we will note that §\ref{subsec:OPSEC4_5} proposes an explication of radio-interferometry as a 
statistical correlation analysis of randomly varying fluxes of hidden photons at high frequencies (not an effect of elusive ``electromagnetic 
waves'' filling full space).
At last in Section \ref{sec:OPSEC5}, we envisage the prevalence of photons at all momenta and their collisions with matter for understanding 
the universe and its spacetime. This prevalence is clear in the $10^{13}$ range at the high end (§\ref{subsec:OPSEC5.1}). We propose here to 
extend it in the $10^{7}$ range of the ZO band, down to RF where photons are known to exist despite beeing hidden individually : indeed they are 
also reflected by mirrors, propagated in vacuum at light speed and absorbed by detectors as light photons are.
At the low end in §\ref{subsec:OPSEC5.2}, even without collecting mirrors, we know that there are photons and we know how to use them 
practically in many applications.

\section{Intensity measurements of light and microwaves (RF)}
\label{sec:OPSEC2}

In the scheme of the International System (SI), the radiative transfer in vacuum from a point source justifies the introduction of a base unit : the 
luminous intensity of light and RF, the Candela\footnote{Candela is the luminous intensity transporting 1/683 W/sr at $\lambda$=0.555 \micro m 
at speed $c$}. Light or RF intensity calibrations\footnote{The NIST microcalorimeter cell calibrates light intensities on the whole visible spectrum 
with a best 0.2\% precision at its center and RF cells are described in \citet{ref:4}} are based on cooled ``microcalorimeters'', that 
measure percents of a Watt, using the temperature raise of a small <<isolated blackbody cell>> inside which the radiation is concentrated 
in order to be ``totally absorbed''. That is a million times more than the power received in a telescope from the brightest star.
For that calibration role in astronomy, we shall use incommensurably more sensitive sensors described hereafter, but all subject to various 
detection limits.

\subsection{measuring light and microwave astronomic intensity time signals} \label{subsec:OPSEC2_1}

The two versions of linear circuits used to amplify, measure and encode an optical signal, both after transforming it into an electrical
signal, are :
\begin{enumerate}
        \item the electrometer for CCD\footnote{or in a Cooled Large Area Photodiode \citep{ref:5}} intensity readout yielding the number of 
electrons integrated during the exposure in each pixel;
        \item the ammeter for a photo-diode or a photo-multiplier (PM) yielding a continuous photo-current sampling record. 
\end{enumerate}
The electrometer yields stable photometric data (cf. §\ref{sec:OPSEC3} and \cite{ref:2}) and the ammeter variable time signals
(cf. §\ref{sec:OPSEC2} and §\ref{sec:OPSEC4})

\begin{figure} \resizebox{\hsize}{!}{\includegraphics{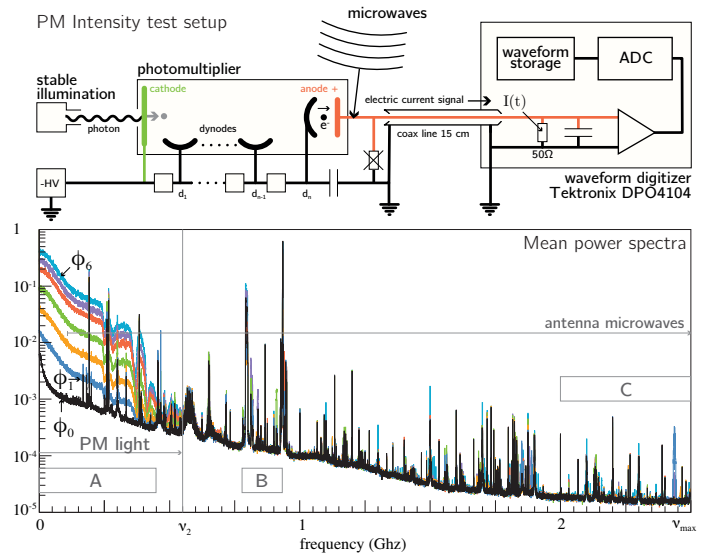}} \caption{\textbf{top)} Setup for recording light and 
microwave intensity $I_k(t)$; \textbf{bottom)} Mean power spectra of $I_k(t)$ obtained by DFT (cf. Fig. \ref{fig:5}) for six light fluxes 
$\phi_k$ falling on PM ($\nu$<$\nu_2$=0.55 Ghz) and a ``dark'' $\phi_0$. Ambient microwaves received by the PM basis up to Nyquist limit 
$\nu_{max}$=2.5 Ghz. Spectra are an average of spectrograms (cf. partial spectrograms of $\phi_5$ : A,C in Fig. \ref{fig:16} and B Fig. 
\ref{fig:15})} \label{fig:1} \end{figure}

We present in Fig.\ref{fig:1} our ammeter setup using a state-of-the-art 5 Ghz/8 bits encoder above the mean power spectra obtained by recording 
seven stable light fluxes analysing them by Fourier transform. It displays not only the light radiant intensity (<0.5 Ghz) absorbed and amplified 
by the PM, but also a forest of RF lines\footnote{which origin is found in §4}, transmitted by the same coax line to a common 50$\Omega$ digitizer.
More precisely, in the PM case the light signal is driven by the emission of a single photo-electron from the cathode at a time $t_k$ which 
produces a cascade ($10^5$-$10^6$) of delayed secondary electrons ejected from the last dynode towards the anode yielding the Single Photon Response 
pulse (SPR), a negative DC current time signal contributing to a global intensity current $I(t)$. Actually a generic representation of such an 
electron current $I(t)$ is found in the Eq. \ref{eq:eq1} from \citet{ref:1} following \citet{ref:6} who considers $K$ thermionic electrons 
emitted in a vacuum tube. Taking the emission of each electron at a random time $t_k$ as a Poisson law process and relying on the principle of 
superposition of electric currents in a circuit, Rice represents $I(t)$ as the sum of identical electron electric impulse responses 
$F(t-t_k)$. This statistical formalism expresses both the DC value of the average current and its root mean square fluctuations. We call it 
classically the ``time series''. Its statistical fluctuations are known as the ``shot noise''.
  \begin{equation}
        I(t)=\sum _{k=1}^K F \left(t-t_k\right)
        \label{eq:eq1}
       \end{equation}

\begin{figure} \resizebox{\hsize}{!}{\includegraphics{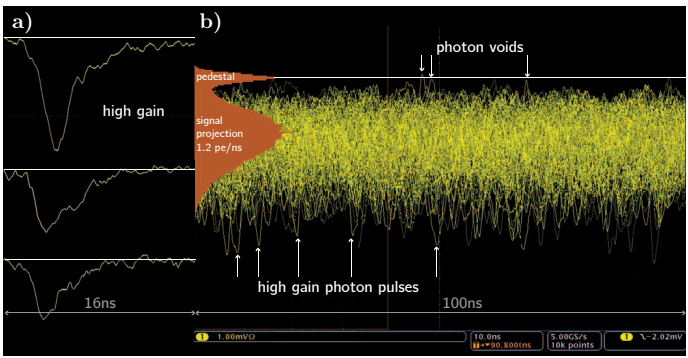}} \caption{PM electric DC signal: \textbf{a)} 3 single photon 
responses (SPR) at the low flux $\phi_1$ (PM digital mode) \textbf{b)} overlapping photons SPR's at a higher flux $\phi_4$ (PM charge 
integration mode) } \label{fig:2} \end{figure}

\begin{figure} \resizebox{\hsize}{!}{\includegraphics{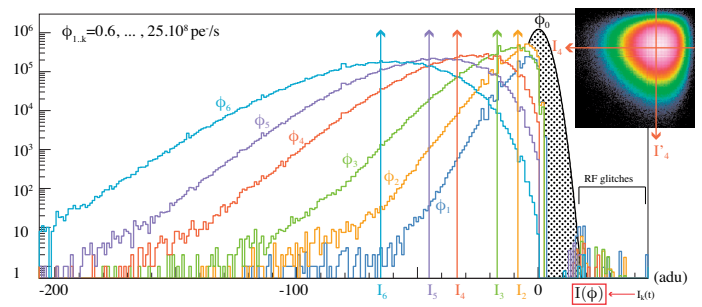}} \caption{Distributions of $10^7$ current samples of the 
pedestal-subtracted PM signals $I_k(t)$-$I_0$ at seven light fluxes $\phi_k$ in a 0 to 2.5 photo-electrons/ns range. A 2-d correlation of two 
different currents at the same flux $\phi_4$ is shown in the insert} \label{fig:3} \end{figure}

We apply this formalism to the 8 bits integer discrete time signals recorded in our PM setup by a 5 Ghz digitizer running continuously 
during 2 ms with a constant Poisson process illumination, at six increasing light flux levels ($\phi_1,...,\phi_6$) from $0.5$ to 
$25\times10^8$ p$e^-$/s, plus two ``darks'' ($\phi_0,\phi'_0=0$). We produce three types of representations of $I(t)$ :

\begin{enumerate} \item Some superimposed oscilloscope traces in Fig. \ref{fig:2}, each representing a continuous signal $I(t)$ obtained by a 
spline interpolation of 5 current samples per ns. \item The seven 256 bin histograms in Fig. \ref{fig:3} containing the $10^7$ integer samples 
(8-bit) of the seven $I_n(t)$ currents recorded at fluxes $\phi_n$. In this plot the samples are considered as successive draws of the random 
variables $I_n$ which probability density law fits the histograms. The correlation analysis of two light signals $I_4(t)$ and $I'_4(t)$ is 
represented in the insert of Fig. \ref{fig:3} by a color encoded density in the plane of the two-dimensional histogram of the random variables 
($I_4$,$I'_4$). \item The six ``mean power spectra'' in Fig. \ref{fig:1} obtained by processing the Digital Fourier Transforms (DFT) of 610 
successive sections ($2^{14}$ samples each) of the $I_n(t)$ signals, as explained later in Section \ref{subsec:OPSEC2_3}. The lower flux 
$\phi_1$ is chosen such that successive SPR in Fig. \ref{fig:2}-a are disconnected\footnote{This is called traditionally the ``digital mode'' 
\citep{ref:8} as opposed to the ``charge integration mode'' of continuously overlapping SPR pulses. PM charge integration was common in the 
nineteen-fifties, before the existence of efficient photodiodes and of transistor amplification to replace photoelectric cells (galvanometers 
directly connected to a photodiode).}. 
\end{enumerate}

The dominant terms in the SPR are the exponential decay time ($RC$<15 ns) and the rise time expressing 
the high frequency cutoff (due to the self in the anode circuit, etc.). However, due to the fluctuations of the electron multiplication by 
each dynode (a factor $\approx$3 following approximately a binomial law) and of the transit time through the 12 dynodes, the SPR is not 
reproducible. The higher fluxes, above $\phi_4$, yield a ``charge integration'' continuous signal shown in Fig. \ref{fig:2}-b by using the 
screen memory of the scope. On the large intensity amplitude side (negative), some rare high gain photon pulses are emerging. On the low 
amplitude side (pedestal), there remain three voids in the distribution of photons where the trace of the scope return to the baseline. 
This test yielded $10^7$ current samples at each flux, which distributions are histogrammed in Fig. \ref{fig:3}. The histogram of the dark 
exposure $\phi_0$ and its gaussian fit centered on the baseline (called ``pedestal'') is cut in half to show its overlap with the light 
exposures histograms. This displays the operating conditions with a PM which gain is a factor ten below what is needed to reach the 
``efficiency plateau'', where the $\phi_1$ and $\phi_0$ histograms dissociate. Measuring the correlation of two simultaneous currents requests 
a 2-d non-gaussian histogram (such as $I_4(t)$ vs $I'_4(t)$ color-encoded in the insert).

\subsection{spectral analysis of a stationary intensity signal}
\label{subsec:OPSEC2_2} 

According to \citet[Section 1.7 p. 306]{ref:1}, the light ``intensity spectrum'' concept was introduced by the famous 
authors\footnote{Rayleigh, Gouy, Plank, Einstein, von Laue and followers} of the quantified black-body radiation theory who wrote a 
theoretical formula representing a global light wave $J(t)$ as a sum of randomly distributed ``optical impulse waves''\footnote{cosine/sine 
waves quantified in a box (not free photon wave packets)} emitted at times $t_k$. Rice points to the analogy of $J(t)$ with Schottky's 
electron current $I(t)$ described in Eq. \ref{eq:eq1} by a time series $t_k$ yielding by Fourier analysis the thermionic intensity spectrum, 
i.e. the shot noise fluctuations around a constant average. The electric impulse response $F(t-t_k)$ in Eq. \ref{eq:eq1} takes the place of 
the optical impulse wave in $J(t)$. Actually what these black-body theorists called a <<light intensity spectrum>>, as reported by Rice, are 
the coefficients $\{a_n,b_n\}$ of a theoretical Fourier series development of their optical wave $J(t)$. This means that $a_n$ and $b_n$ are 
the projections in a vector space of optical waves of infinite duration on a basis of cosine and sine waves at given discrete frequencies that 
yields the light wave $J(t)$ of classical optics when summed. Mathematics tells that this type of discrete Fourier series (DFS) either diverge 
or converge towards a periodic function defined on a finite time interval, but it cannot represent a time series recorded during an 
arbitrarily long period. However we can refer to \citet[][chap.1-3]{ref:9} for relevant mathematical concepts developed in digital signal 
processing, which differ from the functional analysis approach which lead to the inconsistencies mentioned above. In a nutshell : we should 
use discrete Fourier transform (DFT), not DFS, to create a spectral representation of a photon time series. In the following we show how 
splicing a time series of arbitrary length into successive finite series of equal duration $\Delta t$, then applying DFT's to each successive 
splice, yields a series of spectra rigorously equivalent to the initial time series, both being vectors of arbitrary dimension linearly 
related by an orthogonal matrix product :

\begin{itemize} \item[-] in Section \ref{subsec:OPSEC2_2_1}, using a Poisson process random generator, we apply rigorously this mathematical 
method to the ideal case of a stationary photon beam with an impulse response $\delta(t-t_k)$.

\item[-] in Section \ref{subsec:OPSEC2_2_2}, we extend this analysis to the data produced by digitizing the output of a real PM in the charge 
integration mode for six stationary photon beams with various average intensities. The spectral analysis of these data exhibits a perfect 
Shottky behaviour of the anode electron beam, which electric impulse response $F(t-t_k)$ checks exactly the Campbell theorem, i.e. multiply 
the white noise spectrum by the Fourier transform $g(\nu)$ of $F(t-t_k)$. \end{itemize}

\subsubsection{simulated ideal stationary time series} \label{subsec:OPSEC2_2_1}

\begin{figure}
\resizebox{\hsize}{!}{\includegraphics{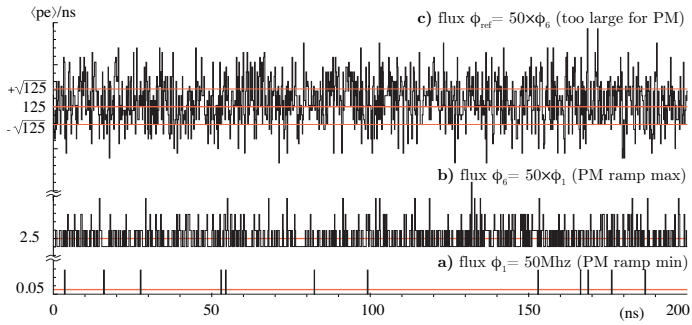}}
\caption{Three randomly generated photon signals sampled at $\delta t=0.2 ns$: \textbf{a)} low flux $\phi_1$ sustainable by a PM in digital mode; 
\textbf{b)} flux $\phi_6$=$50\times\phi_1$ readable by one PM at low gain in charge integration mode; \textbf{c)} high 
flux $\phi_{ref}$=$50\times\phi_6$ measurable by a photodiode or a CCD. AC baselines at 0.05, 2.5 and 125 $\langle pe^-\rangle$/ns are figured 
in red.}
\label{fig:4}
\end{figure}

Ideal photon time series are created by counting the photon time stamps $t_k$ falling in each of the $2\times N$ cells\footnote{adapted to the 
resolution of the time measurement} of width $\delta t$ within a partition of the $\Delta t$ time interval using a binomial\footnote{a 
multinomial law for a photon detector with many channels, typically all CCD pixels in an image and their losses} law random generator. The 
expected value of the photon count in a $\delta t$ interval is a constant and its statistical fluctuation (conditional to a marginal law 
representing the photon source system) is the shot noise. Our practical model was presented in Section \ref{subsec:OPSEC2_2}. It is a PM 
calibration ramp illuminated by a very stable light source at 6 increasing fluxes from $\phi_1$ to $\phi_6$, plus a null flux $\phi_0$. We 
simulated the same ideal calibration ramp numerically. This yielded some stationary photo-currents seen in Fig. \ref{fig:4} with light fluxes 
from $\phi_1$ to $\phi_6=50\times \phi_1$ together with $\phi_{ref}=50\times \phi_6$ identical to the experimental ones in Fig. \ref{fig:3} 
but for an ideal Dirac $\delta(t-t_k)$ impulse response proportional to the photon count. The degradation of this Dirac response by a PM  
is studied in Section \ref{subsec:OPSEC2_2_2}.

\begin{figure}
\resizebox{\hsize}{!}{\includegraphics{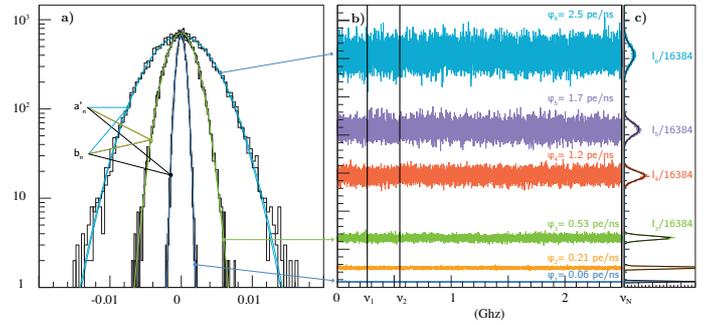}}
\caption{\textbf{a)} distribution of $a'_n$ and $b_n$ (real and imaginary parts of the DFT transform of Poisson signals generated at fluxes 
$\phi_1$, $\phi_3$ and $\phi_6$with Gaussian fits overlaid); \textbf{b)} mean power spectra $\langle p_\nu\rangle_{610}$ ; $\nu_1$ and $\nu_2$ 
are the limits of the real PM spectra in Fig. \ref{fig:6}; \textbf{c)} profile histograms of mean power samples}
\label{fig:5}
\end{figure}

We generate 10 mega-samples records like our experimental test setup does, representing a constant discrete current $I(t)$ holding 2 
ms, which is decomposed at will in a time series of random integer vectors of dimension $2^{m+1}$. Our basic numerical simulation takes 
$m$=13, yielding 610 successive and independent integer vectors of dimension 16384. The measurement time for the $2^{m+1}$ components of a 
single vector is $\Delta t=2^m\times 2 \delta t$. To each vector, we can apply a DFT yielding $N$=$2^m$ pairs of DFT coefficients 
$(a_0,a_N)\bigcup \{a_n,b_n\}_{n=1,N-1}$. The coefficient\footnote{with a global normalization factor $\sqrt{N}$} $a_0$ (null frequency) is 
the total number of photons $K$ counted during $\Delta t$ and the coefficient $a_N$ (N for Nyquist frequency $\nu_N$= $1/2 \delta t$) is the 
difference between the number of photons in either even or odd time samples. The DFT transforms the photon time histogram into two frequency 
histograms $\{a_n,b_n\}_{n=1,N-1}$, with $\nu$=$\nu_N \times n/2^m$. The [1,N] range of index n fixes the frequency range $\nu \in [1/\Delta 
t, 1/2 \delta t] $. Then eventually, for each $\Delta t$ period, we subtract the average current $a_0/N$ (a red line in Fig. \ref{fig:4}) from 
$I(t)$ what transforms $a_n$ into $a'_n$. One advantage of having a realistic statistical scale, a signal/noise ratio below or around 1 (say 
an oversampled signal amplitude) compounded by an enormous number of draws, is to learn how to practice a situation not found in the 
textbooks. Photo-currents generated by the fluxes $\phi_1$ to $\phi_6$, are neither of the Shannon-Nyquist's textbook type (a continuous time 
function with a small gaussian error), nor a decay distribution with an exponential law for $t_{k+1}$-$t_k$. However $(a_n, b_n)$, as an 
orthogonal transform by the DFT matrix of each vector of successive draws of a binomial random variable, are well defined independent random 
variables, linear functions of independent variables. Their 2-d probability distributions are uncorrelated gaussians\footnote{the Fourier 
coefficients $(a_n,b_n)$ are an orthogonal combination of a great number ($2^m$) of independent binomial variables. Both are converging for 
large $m$ towards independent gaussian variables. We have 610 outcomes of these $(a_n,b_n)$ variables, one every 3.2768 \micro s. The 
coefficient $a_n$ is reduced to $a'_n$ ($\langle a'_n\rangle$=$\langle b'_n\rangle$=0)} in the frequency domain (Fig. \ref{fig:5}-a), even at 
low flux when they are sparse in the time domain (Fig. \ref{fig:4}-a). This gaussian property is explained mathematically by statistics 
applied to the photons generated by a Poisson process. The mean square of the intensity fluctuation, computed both in the time and the 
frequency domains, yields identical results as predicted by the Parseval identity. Due to the gaussian property, mean square is easier to 
measure in the frequency domain. The 2-dimensional, centered, uncorrelated gaussian distribution of the bivariate $(a'_n, b_n)$, transforms 
into the product of two 1-dimensional laws : an exponential law for the ``power'' (modulus squared) $p'_n= \rho_n^2 = {a'}^2_n + b^2_n $, plus 
an uniform law for the phase (argument) $\theta_n$.

\setlength{\abovedisplayskip}{1pt}
\setlength{\belowdisplayskip}{1pt}
\begin{equation} 
\begin{split}
        d^2 p'_{\nu }=e^{-(a'^2_\nu + b^2_\nu)/2\sigma^2}\, da_\nu \, db_\nu = e^{-\rho_\nu^{2}/2\sigma^2 }\, \rho_\nu \, 
d\rho'_\nu\,d\theta\\= e^{-p'_\nu /2\sigma^2 } \, dp'_\nu /2\,d\theta
        \label{eq:eq2}
\end{split}
\end{equation}

This transform creates a singularity at $p_n$=$0$ and a cut\footnote{the effect of this singularity is seen in Fig. \ref{fig:6}-b where 
$p_\nu$=$0$ for $\nu$=0.37 and $\nu$=0.41 and, across the cut, the continuity of the phase angles must be restored to express the continuity of 
the source system state variation with time, e.g. in Fig. \ref{fig:18}} at $\theta_n$=$\pm\pi$ . The power spectra of Fig. \ref{fig:5}-b are 
realized by averaging the 610 independent outcomes of  $p_\nu$, for each of the 8192 equally spaced frequencies $\{\nu\}$ from 0.3 Mhz to 2.5 Ghz. 
The mean $\langle p_\nu\rangle_{610}$ of the exponential distribution of power $p_\nu$ is independent of $\nu$ and its root mean 
square is equally $\langle p_\nu\rangle_{610}$. Therefore the average intensity spectra is predicted to be $\langle 
p_\nu\rangle_{610}\pm\langle p_\nu\rangle_{610}$/\mbox{\tiny $\sqrt{610}$}. This is validated in Fig. \ref{fig:5}-b, showing a white noise 
(flat, with a flat gaussian fluctuation).

\subsubsection{experimental PM stationary time series} 
\label{subsec:OPSEC2_2_2}

\begin{figure} 
\resizebox{\hsize}{!}{\includegraphics{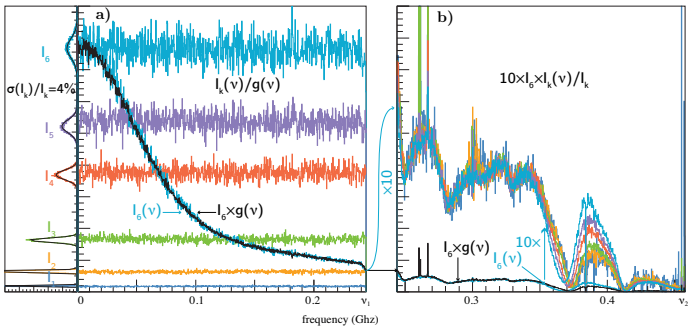}} \caption{The six intensity spectra $I_k(\nu)$ obtained with the PM 
fluxes $\phi_k$ of Fig. 3, are represented in two windows : \textbf{a)} for $\nu_{min}$<$\nu$<$\nu_1$ after flattening (by dividing raw 
spectra $I_k(\nu)$ by $g(\nu)$); \textbf{b)} for $\nu_1$<$\nu$<$\nu_2$, after normalising and zooming (multiplication by $10\times I_6/I_k$). 
The gain curve $I_6\times g(\nu)$ is superposed to the spectrum $I_6(\nu)$ on both windows. Superposition of normalized spectra (right window) 
fails around 0.4Ghz due to the two zeros of $g(\nu)$ interfering with electronic noise subtraction. Raw spectra including noise and parasitic 
microwave intensity are seen in Fig. 1 (PM intensity is null above $\nu_2$).} \label{fig:6} \end{figure}

In the charge integration mode the PM electric signal is not counting photons. As in Eq. \ref{eq:eq1} this is the sum of the electric responses 
$F(t-t_k)$ of the anode to any electron $k$ emitted by the last dynode. This justifies the application of a well known theorem from \citet{ref:7}. 
Accordingly, the Fourier transform $g(\nu)$ of $F(t)$ is factorized in front of the white noise produced by the sum of all individual anode 
electrons phase factors. This fact is illustrated in Fig. \ref{fig:6} by a common gain factor $g(\nu)$ multiplying all PM white noise 
intensity spectra figuring in Fig. \ref{fig:5}-b. The general behaviour of $g(\nu)$ below the frequency $\nu_1$=0.25 Ghz is, as expected, 
a Lorentzian peak of width $\nu$=$1/RC$ ($RC$ is the effective decay constant of $F(t)$). This factor $g(\nu)$ is common to all fluxes, on the 
condition that the intensity spectrum dominates the electronic noise, as it does below $\nu_1$ ($g(\nu_1)/g(0)\approx 4\% $). This model works 
in the $[\phi_1,\phi_6]$ flux range and in the $[0,\nu_1]$ frequency range. Therefore it covers a total intensity range of about 400 (for 
instance in Fig. \ref{fig:6}-b the spectrum $I_1(n)$ is zoomed by a factor 417 to be superposed to $10\times I_6(\nu)$, yielding a precision 
of $\pm 0.1\%$). The electronic noise RMS, measured using three ``dark current'' exposures, is subtracted quadratically from the six intensity 
spectra, for each of the 8192 channels. The fit of these six spectra yields the common gain curve $g(\nu)=\langle I_k(\nu)/I_k\rangle_k$, plus 
one average intensity fluctuation parameter $I_k =\langle I_k(n)/g(\nu)\rangle_n$ per spectrum. The parameters $I_k$ are the abscissa of the 
calibration curve Fig. \ref{fig:7}-b. To sum up, we defined three frequency regions : high frequencies $\nu$>$\nu_2$=0.45 Ghz ($82\%$ of our 
spectra) with no light intensity effect in the PM signal, low frequencies below $\nu_1$ where $g(\nu$) is easy to measure at a $\pm0.1\% $ 
precision and a band around 0.4 Ghz, between the two zeros of $g(\nu)$, which requires a better noise subtraction analysis.

\subsection{auto-calibration of intensity} 
\label{subsec:OPSEC2_3}

The shot noise model of stationary intensity measurement in Eq. \ref{eq:eq1} relates the mean squared current fluctuation 
$\langle(I(t)-\langle I(t)\rangle)^2\rangle$  to the average 
value of $\langle I(t) \rangle$. 

\begin{figure}
\resizebox{\hsize}{!}{\includegraphics{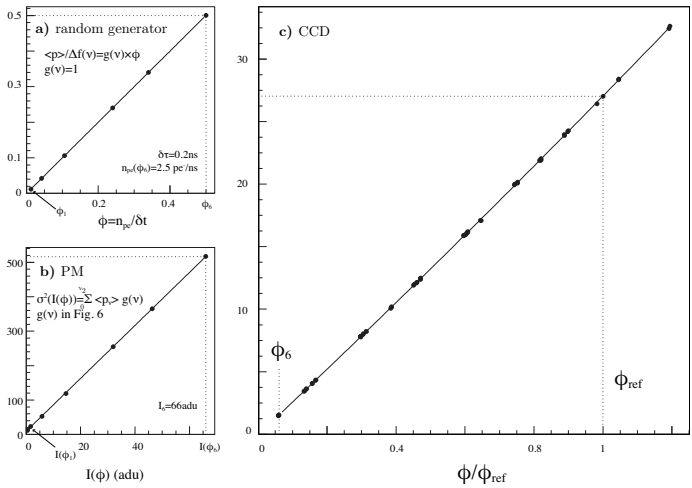}}
\caption{Sum of (flux-fluctuation)\textsuperscript{2} vs flux sum : \textbf{a)} simulated PM data using a random generator sampled at 
$\delta t=0.2 ns$ (channel width $\delta f(\nu)=1/16384$); \textbf{b)} real PM data in charge integration mode (all frequency channels 
summed); \textbf{c)} gain calibration of a Megacam channel ($\langle ke^-/pixel\rangle$ vs relative flux sum).}
\label{fig:7}
\end{figure}

It is a consequence of three joint 
properties : the quantification of intensity, the uniqueness of the electronic impulse response to each single quantum (the ``quantum gain'') 
and the additivity of individual quantum responses which yield a perfectly linear intensity measurement. The result of this model is, as seen 
in Fig. \ref{fig:7}, in three different cases, a precise linear relation between the average current and the sum of its squared fluctuation. 
The slope of this relation calibrate the quantum gain. That common feature is more complex than usually thought. In the bottom left figure (Fig. 
\ref{fig:7}-b), it is obtained using the PM calibration data of Section \ref{subsec:OPSEC2_2_2} at six fluxes $\phi_k$. The abscissa is the 
average photo-current $I_k$ of Fig. \ref{fig:3} and the ordinate is the sum of the square of the intensity fluctuation $\sigma(I_k)$ computed 
by DFT\footnote{taking advantage of the gaussianity of intensity fluctuations in the discrete frequency variable} in Fig. \ref{fig:6}. The 
linear relation is valid experimentally within one per mil. The top left figure (Fig. \ref{fig:7}-a) is obtained by making a DFT of the photon 
arrival times generated by a numerical simulation at same fluxes $\phi_k$ (see Fig. \ref{fig:5}). This is an errorless, purely mathematical, 
result yielding a quantum gain equal to one. It is the mathematical model of a PM operated in the digital mode (and withstanding fluxes above 
$\phi_1$ unlike our PM). The right figure (Fig. \ref{fig:7}-c) is an experimental result obtained independently for each of the 72 CCD 
channels, by summing 4.7 megapixels content (\cite{ref:2}, Fig. 20) at 70 different fluxes up to $\phi_{ref}$=$50\times \phi_6$. It yields a 
reproducible, measurable response for each photo-electron detected in each pixel. This voltage response $e/C_{in}$, amplified by the CCD 
channel amplifier common to all pixels, yields the quantum gain \footnote{effect of one photo-electron charge on the $C_{in}$$\approx$50 fF input 
capacitance of the channel amplifier}. Joining the data of all 72 CCD channels brings the overall statistical precision of this measurement to 
$\approx$1 ppm for the $\phi_{ref}$ flux, just limited by the statistical error due to the number of photo-electrons analysed ($5\times 
10^{12}$). There is a mismatch between this excellent self-calibration of the number of photo-electron counted by the CCD and the 
determination of the incoming photon number, expressed by the quantum efficiency parameter $\epsilon_k$. Its complement 1-$\epsilon_k$ counts 
the number of photons reflected on the windows or thermalized locally in the CCD (eventually after emission of a lower energy photon by local 
electron-hole recombination). Last refinement, the setup permits to equalize independently the quantum efficiency of different CCDs and LEDs 
and to test their dependence on external parameters such as temperature. Returning to the PM, the single electron impulse response given by 
Eq. \ref{eq:eq1} is the same whether in charge integration mode or in single photon mode. It is applied to each anode electron detected at 
time $t_k$ with a response $F(t-t_k)$ obtained by back-DFT transforming $g(\nu)$ into $F(t)$. Practically, that means that one obtains the 
SPRs seen in Fig. \ref{fig:2}-a by folding the cathode-anode transit time distribution of the $\approx$$10^5$ anode electrons with the single 
photon response $F(t-t_k)$ (taking the photon arrival time on cathode as origin of time). Therefore the quantum gain defined by the slope of 
the Fig. \ref{fig:7}-b is an artifact, not the response to one photo-electron. It represents the average power dissipated by the average 
current traversing the 50$\ohm$ resistor which terminates the coaxial line. This current is produced by the cascades of secondary electrons 
initiated successively by each photo-electron created on the cathode during 2 ms.

\subsection{the intensity spectrum of stationary light}
\label{subsec:OPSEC2_4}

We have replaced the ambiguous light ``intensity spectrum'' concept of \citet{ref:1} by a rigorous definition supported by most precise 
practical examples : the counting of the photon numbers (photon-resolved signal sampling) detected in successive $\delta t$ intervals 
(discrete time signal) and of the digital signal processing mathematics applied to these signals. The first step of processing is the 
application of the DFT matrix to a series of integer vectors (finite time series of 2$N$ samples), which yields for each vector two 
orthogonal\footnote{commonly said in quadrature} amplitude spectra $\{a_n\}$ and $\{b_n\}$ of dimension $N$ or equivalently the two other 
compounded spectra, the power $\{p_n=a^2_n + b^2_n \}$ and the phase $\{\phi_n=\arctan{(b_n/a_n)}\}$ . This series of spectra is a lossless, 
linear and reversible representation of the original finite time series. The second step consists of two statistical correlation analyses 
using linear regressions of, either intensity spectra or time series vectors. Both analyses are algebraically transposed by DFT, yielding the 
Parseval identity. For a Poisson emission process, we validated the approach\footnote{the repetition in time is regarded as independent draws 
of the random spectral vectors $a_n$ and $b_n$} mentioned by Rice both mathematically and experimentally by a rigorous statistical analysis of 
the averages and the RMS of the series of random vectors $p_n$ and $\phi_n$ obtained from time series containing a large number of particles 
emitted by a common source. The spectral analysis of stationary processes yielded a very compact statistical representation : a constant 
average intensity spectrum which statistical fluctuations are described either by a ``white noise'' (Fig. \ref{fig:5}) or by a ``grey noise'' 
(Fig. \ref{fig:6}) when taking into account the non-Dirac electric response $F(t-t_k)$. This grey noise exemple from our PM test receiving 
stationary light fluxes yielded the average power spectra in Fig. \ref{fig:1}. Below $\nu_2$=0.45 Ghz, it is dominated by the sum of the power 
spectra of all the single anode electron responses $g(\nu)$ ($\approx 10^5 \times g(\nu)$ for the sum of all electrons created by one 
photo-electron). The electric response to light $g(\nu)$ being null for $\nu$>$\nu_2$, this part of the intensity spectrum in Fig. \ref{fig:1} 
is reserved to microwaves (cf. Section \ref{subsec:OPSEC4_3_1}).

This mathematically correct definition of stationary intensity spectrum is an important step before studying variable intensities in the next section.

\section{Zukunftoptik}
\label{sec:OPSEC3}

In 1932 Born, while he was contributing to the birth of quantum mechanics for which he received the Nobel prize in 1954, asserted in 
\citet{ref:10} that ``conventional optics'' had to be replaced by ``Zukunftoptik''(ZO), alias Future optics. We guess that Born's motivations 
were : first that the photon had just been discovered experimentally and second that his own scattering theory applied to photon as it is in 
Section \ref{subsec:OPSEC3_1} could explain the fundamental quantum bases of optics. We have shown in \cite{ref:2} and will repeat more 
completly here that this is the only way compatible with experimental facts. Born introduced in addition his mathematical approximation 
techniques very popular among theoreticians for its first applications to quantum mechanics. But, except for a few simple cases (planar or 
spherical optical potential), Born QM scattering on complex geometrical objects lead to intractable mathematical problems and therefore photon 
ray tracing approximation remained the practical solution used by astronomy.

But fundamentally ZO, contrary to conventional optics, distinguishes the quantum signals conveyed by the propagation of each individual 
photon wavefunction which yield the ``optical spectrum'' from the classical signal produced by the variation in time of a photon source 
emission process which yields the ``intensity spectrum''. The explanation of the optical spectrum is offered by Born's scattering of single 
photons on the optical potential of whole condensed objects such as gratings, prisms, atoms or molecules, droplets, defects of mirrors and of
lens. Our experimental data, represented with a high precision by the PQM model\footnote{model overly simplified since photon spin is ignored, as 
done by Born scattering}, imposed a very complex submicron representation of object interfaces and of their 
constituting media including their defects. This result was obtained using a huge computing power to process high photon statistics and an 
extra-sensitive CCD camera yielding photon-resolved intensity signals both available only for the last decade. We recall in Section 
\ref{subsec:OPSEC3_1} how light ZO can use these progresses. The main feature, presented in Section \ref{subsec:OPSEC3_2}, is a separation of 
angular and time (or time of flight) variables. A discontinuous photon emission process added to a stationary background and to electronic 
noise in the photocurrent yields a very complex time signal. It is usually called a ``random noise'' signal. Additional informations are 
needed to extract the genuine intensity variation from the electronic noise within a random noise signal. They are found in a statistical 
correlation analysis of duplicated intensity signals : genuine intensity variations are duplicated but not the Poisson shot noise.

Owing to quantum mechanics, the duplication of the wavefunction emission, its free propagation in vacuum and its elastic scattering on 
condensed matter can be as perfect as the duplication of wave functions inside indiscernible atomic or molecular bound states. Using the data 
taken from this exemple, we apply in Section \ref{subsec:OPSEC3_3} a classical regression analysis to test the conformity of the two time 
series produced by two duplicated photon-resolved signals which contain a common 3.5\% rms random intensity variation superposed to a 
stationary process. It aims to give an insight into the statistical signal processing methods applicable to a large number of resolved 
photons (between $10^9$ and $10^{14}$ in our work). In Section \ref{subsec:OPSEC3_4} these methods are transported by DFT into the discrete frequency 
space. This yields clearly the ``intensity interference'' effect. We observe in Section \ref{subsec:OPSEC3_5} that the conclusion of our 
experimental analysis is conform to Dirac quantum mechanics theoretical principles. We did not develop the fact, implied in Dirac statements, 
that data and quantum mechanical theory are totally incompatible with Maxwell's <<Electromagnetic Theory of Light>> 
(ETL)\footnote{\citet{Max1865}}.

\subsection{Born's elastic scattering of single photons on optical surfaces} \label{subsec:OPSEC3_1}

\begin{figure*}
    \centering
    \includegraphics[width=17cm]{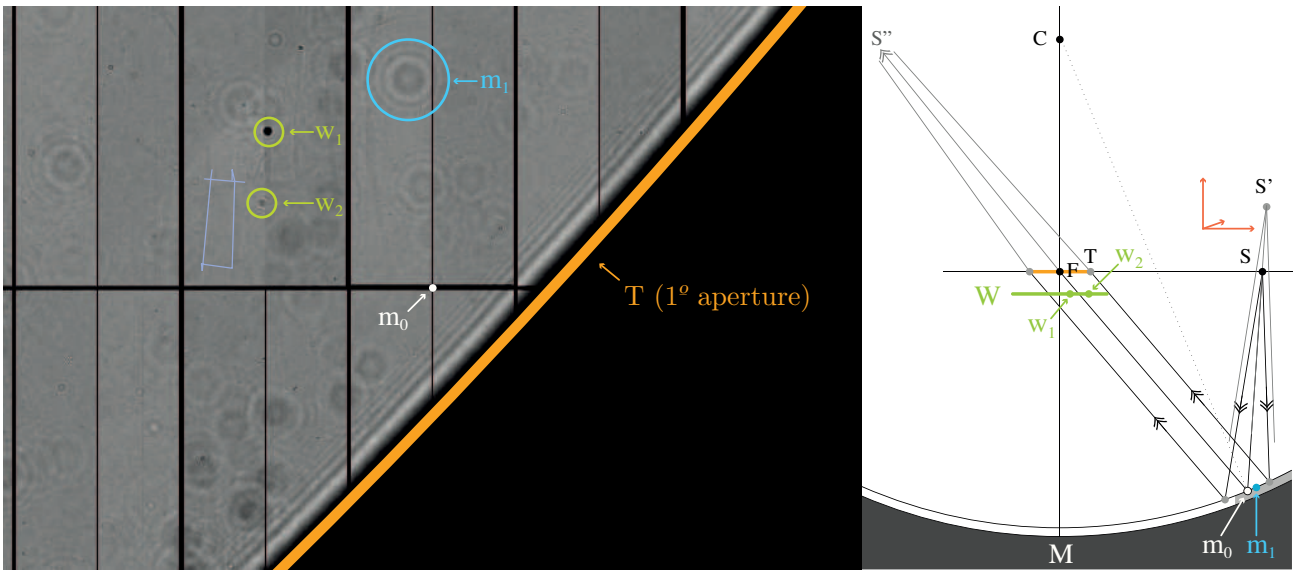}
   \caption{LED photons emitted by Sndice source S' are counted by CCD pixels in the Megacam camera around the primary focus F, after Born's 
elastic scattering of photons on the CFHT telescope mirror M and entering the camera through window W and lens $W_i$. The resulting image 
visualize the variation of photon count around its mean by a linear gray scale (black-white $\leftrightarrow \pm 5\%$).
    It shows sequences of small rings caused by point defects ($m_i$ for mirror defect; $w_i$ for window defect) and of large diffraction 
rings at the transition T between the mirror and its dark support (a $\Phi3.6m$ orange circonference). 
Spacings between these rings, are given approximately by the Cornu spiral, which characterize the distance between scatterer and detector 
($FM=15m, FW<30cm$)}
    \label{fig:8}
\end{figure*}

The ZO refoundation of conventional optics is based on Born's scattering concept applied to light : the quantum mechanical elastic scattering 
of spinless photons on macroscopic condensed objects (represented by an ``optical potential''). We illustrate this concept and its limits in 
Fig. \ref{fig:8}, using the experimental optical effects of Born scattering on a telescope mirror M or through a cryostat window W and a few 
lens $w_i$, which have been measured very precisely\footnote{at its $10^{-6}$ quantum statistical precision 
limit} by our experimental analysis. This analysis used the following means : 
\begin{enumerate}
    \item a test bench characterization of the conical beam emitted by the SNDICE source S in free space; 
    \item a 3-d displacement of the source parallel to each axes of the telescope mirror using the telescope motion control; 
    \item a variation of the source intensity using electronic controls; 
    \item repeated stable exposures with variable durations fixed by LED shutters (either mechanical or electronic). 
\end{enumerate} 
In Fig. \ref{fig:8}, the source S in the focal plane has been moved to S' (SS'=5 m), parallel to the optical axis, what changes the reflected 
photon beam from parallel to conical, converging towards an image S'' of S' at 60 m above the mirror. The beam cross-section in the focal plane 
stays contained in the aperture (radius FT) of the camera, which contains the $\pm1^\circ$ aperture of the SNDICE source. The stray light 
falling in the CCD detector outside the FT circle edge is minimal. We recognize in this example several characteristic features of ZO affecting 
either intensity or wave function signals. Direct illumination is dominant, but we perceive also secondary illumination caused 
by back-reflection of light on the CCD surface, followed by a forward reflection on a lens surface (with a pincushion distortion enhanced 
by violet lines) and then a detection by CCDs superposed to the direct illumination image.

\subsection{duplication of intensity time signals}
\label{subsec:OPSEC3_2}

The conditions for creating some optical signal duplicates are usually called partial coherence : a variable photon emission process from a 
point source illuminates various solid angles which are subtended by independent detectors. If the detectors illumination is direct, that is 
only by propagation in vacuum, a photon has a probability $\delta\phi=\phi(t,x,y)\delta(t,x,y)$ to be detected at a given time and space 
location, which conveys an intensity signal varying in 3 dimensions. Following this ``direct illumination'' model, we represent $\delta\phi$ 
as the product of : \begin{enumerate}
    \item a time sampling factor, the linear flux $\phi(t)\delta(t)$
    \item a constant normalized (i.e. unitary) x-y position sampling factor $A_k=|\psi(x,y)|^2\delta(x,y)$
    \item a quantum efficiency constant $\epsilon_k$<1 which, multiplied to $A_k$, yields the probability to count a photo-electron in a 
$\delta(x,y)\delta(t)$ cell. \end{enumerate}

In short, time and position variables are separated. Consequently the position variation and the time variation of intensity signals define 
two independent random variables. The former is defined by the wave-function of a photon emitted by the point source, propagated in vacuum and 
even (as in Section \ref{subsec:OPSEC3_1}) elastically scattered on optical surfaces. The latter is defined classically as the ``luminous flux 
$\phi_v$'' of the source (a SI standard definition) or it could be defined as the photon flux. Due to the separability 
of its time and its position variations, intensity (say the photon number detected in a spacetime cell) can be integrated separately either in 
time, or in space inside the detector cells. The total number of these intensity cells and their dynamical range are a prime factor of cost 
and of dissipation of the electric power used for amplification. Therefore it makes sense to conceive the photon counters for large statistics 
experiments either as a time-like or as a space-like signal analyser and to build the detector system accordingly. Our example of a space-like 
photon-resolved signal analyser for up to $10^{14}$ photons, is Megacam. The intensity is integrated in time 
during a frame exposure (0.001-100 s) and distributed in space over a 2-d grid of $3.5\times10^8$ pixels. The field of pixel counts $A_k(x,y)$ 
can be Fourier transformed into a space-frequency 2-d Fresnel spectrum which measures precisely the kernel of the Fresnel diffraction integral and
yields the transform of the autocorrelation of both image contents. All the operations that we have described for two $x$-$y$ intensity image 
variation analysers (2d spectra and image correlation) are analog to what would be done for two time-like intensity signal variation analysers
(1d spectra and signals correlation). However we do not dispose of sufficiently fast photon-resolved sampling devices to match the 5 Ghz
frequency of our current sampling that we do not control or understand at single photon level. 
Therefore we simulated photon-resolved time signals either with a random photon generator as in Fig. \ref{fig:4}, or in the next subsection 
by using duplicated series of photon samples accumulated experimentally in a given Megacam CCD column ($y$ coordinate photon-resolved intensity 
signals) within a succession of image duplicates, as if they were the time series from different detectors illuminated by the same point source.

\subsection{cross correlation of photon-resolved time signals duplicates}
\label{subsec:OPSEC3_3}

\begin{figure} \resizebox{\hsize}{!}{\includegraphics{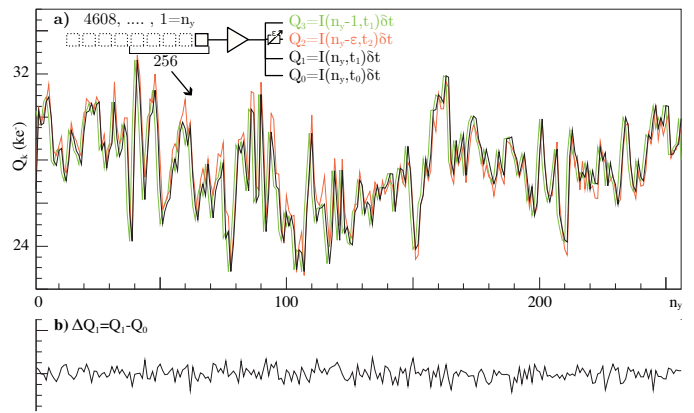}} \caption{\textbf{a)} Four 'random' optical signals $Q_i$ from same 
column in 25 CCD exposures duplicated at times $t_i$. They are converted into photo-electron counts with a $\approx0.01\%$ precision. CCD 
signals at $t_0$ and $t_1$ being perfectly identical, $Q_0$ and $Q_1$ are superimposed. \textbf{b)} The difference $\Delta Q_1=Q_1-Q_0$, has a 
pure gaussian statistical fluctuation with a null expectation value.} \label{fig:9} \end{figure}

\begin{figure} \resizebox{\hsize}{!}{\includegraphics{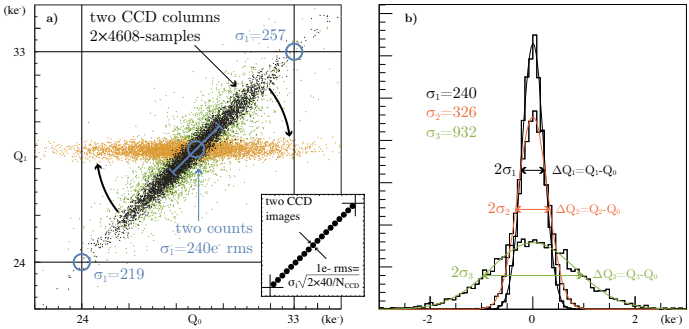}} \caption{ \textbf{a)} Linear regression of $Q_1$ (or $Q_3$) vs 
$Q_0$ : correlation plot of the 4608 pairs of photon counts falling in the same channel figured as black (or green) points. Circles shows the 
$\sigma_1=\sqrt{Q_0}+Q_1$ rms limit of a pair [$Q_0$ ,$Q_1$] of photon counts ($\langle Q_0\rangle$=28.8 K$e^-$). Orange points show the 
rotation of ($Q_0$,$Q_1$) to its principal components axis. Insert shows the effect of integrating all the $N_{CCD}$ pixels into 40 intensity 
bins, increasing the statistical precision by 3 orders of magnitude. \textbf{b)} Increases of $\Delta Q_2$ and $\Delta Q_3$ compared to 
$\Delta Q_1$ are due to the translation of $Q_2[1\mu]$ and $Q_3[15\mu]$ ($\sigma_1 \rightarrow \sigma_2$ and $\sigma_3$).} \label{fig:10} 
\end{figure}

We built the discrete intensity photon-resolved time signals $Q_i(t)$ as described above, to test our cross correlation algorithms. Four 
waveforms $Q_0$ to $Q_3$ are presented in Fig. \ref{fig:9}, in a digital scope format with an integer amplitude scale graduated in 
photo-electron number (not in ADC units) because they are photon-resolved. We recall that these signals conforms strictly to the PQM model in 
which a point source broadcasts perfectly duplicated intensity variations signals in all directly illuminated detectors and respects the 
multinomial probability law describing the various photon count fluctuations. The instrumental problems have been identified and mitigated. 
The charge sampling has a proven 1 ppm photon counting statistical precision. When we interpret $n_y$ CCD samplings as successive digital 
scope 0.2 ns samplings, the average CCD charge sample $\langle Q(t)\rangle$=28.8 k$e^-$ corresponds to a 200$\times\phi_{ref}$ PM flux, 
what is 16 \micro W, a million times higher than the maximal flux $\phi_1$ accepted by a PM in a photon counting mode and thousand times 
smaller than a standard NIST thermal intensity measurement. The $\pm$1.5 k$e^-$ variation range of the $Q_i(t)$ signals ($\approx$ 3.5\% 
rms of total current) can be represented by a ``random signal'' added to a stationary background. The four intensity signals $Q_0$ to $Q_3$ 
are taken from three Megacam images among the 25 recorded within the same hour, at times $t_0$,$t_1$,$t_2$. Two signals $Q_0$ and $Q_1$, 
being successive exposures ($2'$ delay), are perfect duplicates but the third one $Q_2$ taken one hour later is translated by $\epsilon$=1/15 
pixel-width (i.e. $\epsilon$=1 \micro m) along the $y$ axis, due to a 8.0 nrad angular drift of the LED source center in the telescope 
reference frame. The last one $Q_3$ is identical to $Q_1$ but artificially translated by a full pixel width $\epsilon$=15 \micro m, owing to 
a pixel number shift ($n_y\rightarrow n_y-1$). The statistical analysis which produced these data, tested the sameness of the time-signals 
$Q_0$ to $Q_3$ by a regression analysis which is represented in short in Fig. \ref{fig:10}. The fine tuned analysis\footnote{details 
found in the report \citet{ref:14}} is much more complex than what is evoked here. It takes into account many details, 
ultimately validated by the narrowing of the distribution of difference current $\Delta Q_i$=$Q_i-Q_0$. The minimal RMS 
width $2\sigma_1$ of this distribution seen in Fig. \ref{fig:10}-b is fixed by the statistical fluctuation of each difference of photon 
counts duplicates and validated by the white noise nature of $\Delta Q_1(t)$ in Fig. \ref{fig:9}-b. It provides the quantum gain, hence the 
absolute photometric calibration of the Fig. \ref{fig:7}-c, obtained after correcting the various side effects. Let us mention the main 
corrections :

\begin{itemize}
    \item[-] optimisation of the statistical estimation method for the $Q_i\otimes Q_j$ non gaussianity ($\sigma_j$ varies with $Q_j$); 
    \item[-] optimisation of the CCD photo-electric charge collection (parallel light ray incidence, photodiode bias voltage adjusted, charge 
transfer efficiency measured);
    \item[-] check of the linearity of the charge versus integrated flux relation; 
    \item[-] calibration of the relative response of different CCD channels and mitigation of some electronics defects (called ``system noise''); 
    \item[-] photometric effect of diffraction, omitted by ray optics but predicted by the Fresnel diffraction integral and seen in Fig. 
\ref{fig:8}, explained by Born as the elastic scattering of photons on the mirror optical potential at the edges of the light field (Cornu 
spiral) or on point defects of the mirror (Strehl ratio);
    \item[-] translation in the $y$ dimension of the intensity signal with an effect on duplication seen in  Fig. \ref{fig:10}-b.
\end{itemize}

The detection and the correction of this translation effect was done by extending the statistics from the number of photons in one pair of 
pixels ($\approx$2$\times$28800), to all pairs of pixels in a given CCD channel (i.e. $N_{CCD}$=4.7$\times10^6$) and then to the 72 channels 
in one CCD image. In turn this high statistical precision allowed a fine modeling of the effect and yielded an exceedingly good sensitivity to 
translation. All this was made possible by the verification of the hypotheses pertaining to the PQM model and to the mitigation of megacam 
electronics (such as the linearity of the CCD response to the intensity variations in the full range of Fig. \ref{fig:7}-c or in the limited 
range of Fig. \ref{fig:10}-a). It was essentially due to the perfect duplication of the spatial frequency intensity variation (3.5\% rms) of the 
photon resolved signal above a plain uncorrelated Poisson background. We recall that our algorithm measuring the translation of an image 
respectively to another along the $y$ (or the $x$) coordinate is based on the image-to-image correlation of one of the four correlation 
estimators linking two pixels intensities among a two-by-two matrix of neighbor pixels \citep{ref:14} \footnote{in image processing 
language they are called ``partial derivative equation'' (PDE) filters, by reference to their relation with the differential operators 
defining the partial derivatives of a continuous 2-d function}. This algorithm measures a 2-d translation vector between both image fields as 
a whole, with a precision three orders of magnitude better than the intrinsic $x$-$y$ resolution of the image (the pixel width). This well 
understood statistical correlation of two duplicated intensity variation signals either in the time or in the space domains can be transported 
by DFT in their corresponding frequency domains. We show in the next paragraph how this frequency correlation of two intensity time signals 
duplicates is the essence of the ``intensity interference'' effect.

\subsection{correlation between varying intensity signal duplicates in the DFT frequency space}
\label{subsec:OPSEC3_4}

\begin{figure} \resizebox{\hsize}{!}{\includegraphics{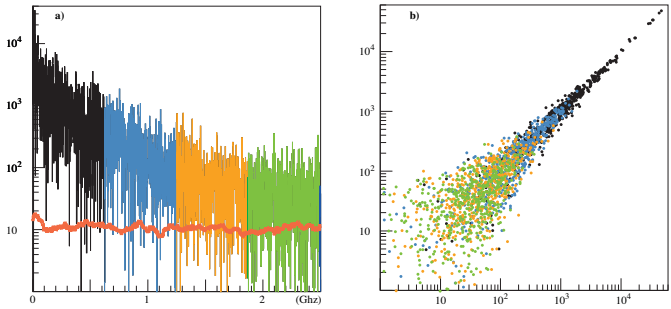}} \caption{\textbf{a)} Power spectra of $p_n(Q_0)$ and $p_n(Q_1)$ 
(four colors) plus $\Delta Q_{1-0}$ a pure photon shot noise (overlaid orange). \textbf{b)} correlation of power spectral densities $p_n(Q_0)$ 
vs $p_n(Q_1)$ (2047 frequency channels).} \label{fig:11} \end{figure}

In the general case of the ``random noise'' emission processes such as the $Q_i(t)$ signals seen in Fig. \ref{fig:9}-a, the spectra, like the 
$Q_0(t)$ power spectrum shown in Fig. \ref{fig:11}-a, may decrease by three orders of magnitude, while their shot noise difference $\Delta 
Q_1(t)$=$Q_1-Q_0$ shown in Fig. \ref{fig:9}-b has a flat intensity spectrum (orange in Fig. \ref{fig:11}-a). The result is a 
signal/noise ratio very small in the higher frequency quarter (black) of the intensity spectrum. Let us remind that the flat shot 
noise spectrum in our exemple is due to the 96.5\% stationary process contribution to the total photon count, while the decreasing spectrum is 
only due to the 3.5\% power of the variable processes. Putting together the $p_n$ and the $\phi_n$ spectra computed using a common discrete 
time series $Q(t)$ of duration $\Delta$t permits a complete recovery of $Q(t)$ by using a discrete Fourier back transform 
(DFT\textsuperscript{-1}). This statement comes with two caveat's.

\emph{First caveat} : the two spectral coefficients $a_0$ and $a_N$ do not figure as such in the intensity spectrum but they are necessary to 
recover $Q(t)$. We treat them apart. The other coefficients $\{a_n,b_n\}_{n=1,N-1}$ encode the intensity fluctuations, not the absolute 
intensity, say for electronicians they represent an AC signal instead of a DC signal. The power variable $p_n$ is an unusual type of random 
variable analog to a 2-d $\chi^2 $ variable with an exponential probability density restricted to the positive half-axis, what creates a 
singularity at origin already discussed with Eq. \ref{eq:eq2}. Therefore the power spectrum in a single channel has little meaning by itself, 
but it could be cross-compared with the neighbouring frequency channels or auto-compared with the next spectrum from a continuous spectral 
time series. In order to process a long time series, one creates a partition of this series into adjacents time segments of length $\Delta 
t$=$2N$$\times$$\delta t$ and transforms each segment by DFT, to yield a 2-d representation of the data which encodes a time series of 
arbitrary length. This time/frequency representation will be rendered in the following by 2-d ``spectrogram'' images one for $p_n$ and one for 
$\phi_n$ (such as in Fig. \ref{fig:15} or Fig. \ref{fig:16}). It comes in addition of the intensity curve $a_0(k)$ which counts the number of 
photons detected in the $\Delta t$ segment number $k$. The number $k$ is the discrete position of the segment in the time series. It is the 
discrete time variable defining $a_n$ ``intermediate frequency'' corresponding to each line of the 2-d spectrogram.

\begin{figure}
\resizebox{\hsize}{!}{\includegraphics{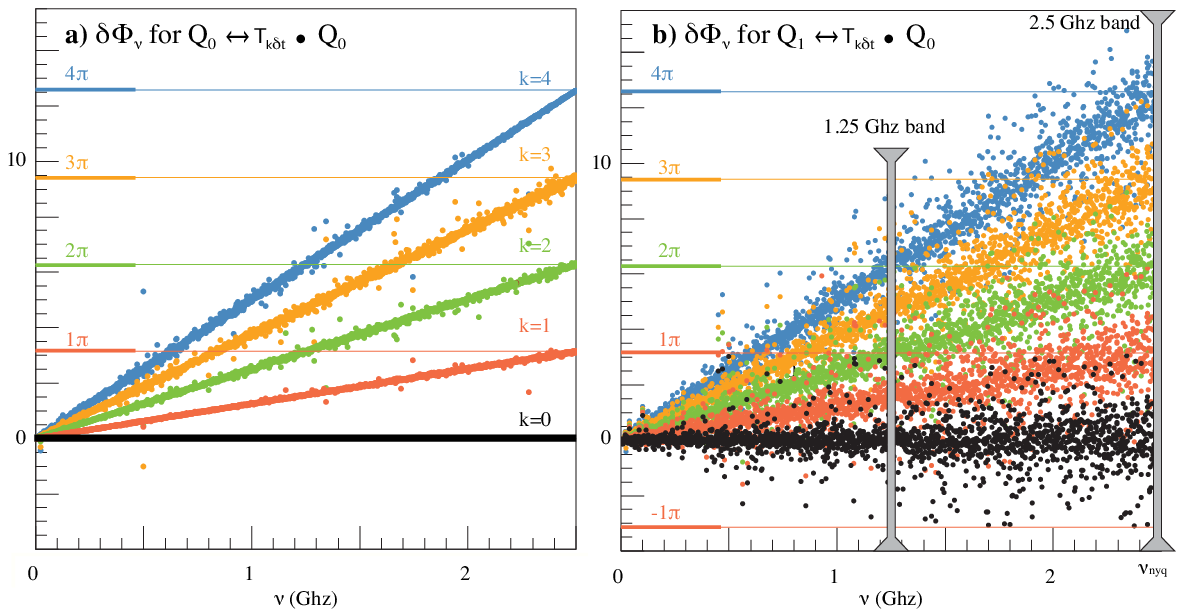}}
\caption{Phase difference $\delta \phi_\nu$ vs frequency $\nu(n)$ ($\nu=\nu_{nyq}\times n_{channel}/2048$) : \textbf{a)} For a 
$k \delta t$ time shift ($k=0,..,4 ; Q_0\rightarrow T_{k\delta t}\bullet Q_0$); \textbf{b)} For a $k\delta t$ time shift plus a duplication 
($Q_0\rightarrow Q_1$ bringing shot noise).}
\label{fig:12}
\end{figure}

\begin{figure} 
    \resizebox{\hsize}{!}{\includegraphics{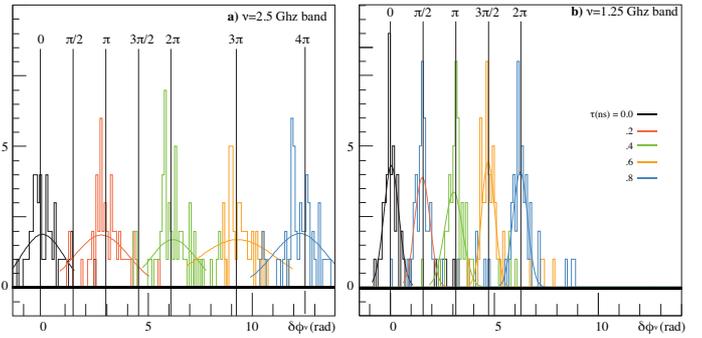}} \caption{Histograms of the $Q_0(t)-Q_1(t)$ phase difference 
$\delta \phi_\nu$ (due to various time shifts $\tau$) in two $\nu$ bands of equal width (marked by gray vertical bands in Fig. 12-b) : 
\textbf{a)} around 2.5 Ghz; \textbf{b)} around 1.25 Ghz} \label{fig:13} \end{figure}

\emph{Second caveat} : When we sample a time signal using a single detector in step with its own local clock, it yields an intensity time 
series $Q(t)$. Then the measurement of the phase $\phi_n$ is not ambiguous. Therefore one can back-transform phase into local time or compare 
the relative timing of two consecutive events recorded by the same clock. However when one emission event in a given source generates 
duplicated signals in detectors A and B, each using their own local clocks, the source-to-detector propagation generate two different ``time 
of flight'' (TOF). The two signals $Q_A(t)$ and $Q_B(t)$ are built with an unknown difference of ``time of flight'' between them. This ``phase 
ambiguity'', contrary to the quantum wavefunction phase ambiguity\footnote{a Barrelet's ambiguity in the analysis of particle scattering 
amplitudes}, can be cured by the determination of the TOF difference $t$ resulting in a translation in time $T_\tau$ such that $Q_B$=$T_\tau 
\bullet Q_A$, as demonstrated hereafter using a cross-correlation test. The conformity of two duplicate intensity signals $Q_A(t)$ and 
$Q_B(t)$ (in our exemple the conformity of $Q_0$ and $Q_1$ duplicates is perfect within a 1 ppm precision) must be tested both in the 
amplitude and either in the time or in the frequency representation. The amplitude conformity test in the time representation was recalled in 
the regression plot of Fig. \ref{fig:10}. Its counterpart in the associated frequency representation is the regression plot of the ``power 
samples'' $p_\nu(Q_1)$ and $p_\nu(Q_0)$ (cf.  Fig. \ref{fig:11}-b). The time conformity test in the time representation is mainly the 
comparison of the dispersions $\sigma_i$ of the difference of the amplitude of duplicated signals done in Fig. \ref{fig:10}-b. This test is 
blind, because it does not measure directly the main cause of dispersion : the TOF difference $\tau(k)$=$k\delta t$. In the frequency 
representation $\tau(k)$ is measured directly in Fig.  \ref{fig:12}. by the slope of the relation between the phases $\phi_n(T_{k \delta t} 
\bullet Q_0)$ and $\phi_n(Q_1)$ (TOF difference is determined by this linear regression of phase, while gain is determined by a regression of 
amplitude). This slope is rigorously null (black points in Fig. \ref{fig:12}-b) for perfect duplicates $Q_0$ and $Q_1$ with $\tau$=0. For 
imperfect duplicates like $Q_0$ and $Q_2$ the corresponding regression slope is significantly positive and compatible with the value 
$\tau$=$\delta t/15$ quoted in Fig. \ref{fig:10}-b, which is determined by a much larger statistics. In order to check and to calibrate this 
conformity check, we applied it in Fig. \ref{fig:12} to the signals obtained by shifting the discrete integer samples of $Q(t)$ by a few units 
($k$=1 to 4). The effect of this translation $Q(t)\rightarrow T_{k\delta t}\bullet Q(t)$ is represented by a linear phase shift $\delta 
\phi_\nu=k\pi\times \nu/\nu_{nyquist}$, which is verified by a linear fit to the regression plots in Fig. \ref{fig:12}-a and Fig. 
\ref{fig:12}-b ($\delta t$ is the sampling period yielding $\nu_{nyquist}$=1/2$\delta t$). This relation is analog to the well known 
expression of the composition of Fourier transform and translation resulting in a product by a complex exponential with a linear phase factor. 
In our algebraic context of discrete variables and of DFT matrices applied to finite time series, this linear phase-frequency relation is a 
good approximation with a superior bound of order $k/2N$ expressing the effect of the subset of samples not conserved in the translation. This 
``edge'' effect is seen in Fig. \ref{fig:12}-a as a residual dispersion of points around the straight line fits. The overall effect of 
duplication, in which the photon shot noise is joined to the time shift, is seen in Fig. \ref{fig:12}-b. It is dominated by the photon shot 
noise, particularly at high frequency, because the statistical weight of the same bandwidth at different frequencies expressed by the 
intensity spectrum is a factor five larger at 1.25 Ghz than at 2.5 Ghz and a factor thousand larger at 0.1 Ghz.
 Fig. \ref{fig:13} represents the spread of the phase samples due to the shot noise in the 1.25 and the 2.5 Ghz bands, as extracted from Fig. 
\ref{fig:12}-b. The parameter relevant for what will be called intensity interference is the difference of TOF $\tau$ between a common source 
and the two detectors yielding the $Q_0(t)$ and $Q_1(t)$ signals. In this exemple the five values of $\tau$ from 0 to 0.8 ns are color coded. 
The comparison of the measurements of $\tau$ at 1.25 and 2.5 Ghz shows a better time resolution at the lower frequency value, although the 
wavelength associated with the lower frequency is larger. The overall time resolution is obtained in the time representation of the $Q_0(t)$ 
and $Q_1(t)$ signals in Fig. \ref{fig:9} by minimizing their $\chi^2$ distance which is precisely estimated by a gaussian law (cf.  Fig. 
\ref{fig:10}-b). In the frequency representation we do not have laws to represent the $N$=2048 variables $\delta \phi_n$, which are necessary 
to fit a straight line on $\delta \phi_n$ vs $n$. But empirically we have estimated a different gaussian law for each narrow band (grouping 40 
$\delta \phi_n$) among the 51 composing the spectrum (as suggested by Fig. \ref{fig:13} for the mid-frequency and highest frequency bands). 
Fitting these gaussians on each $Q(t)$ signal produced a coarse spectrum of 51 points from which a linear fit yielded an approximation of the 
time shift $\tau$ between the $Q_0(t)$ and $Q(t)$ signals with a precision of $\sigma_\tau$=0.01 ns rms (for $\delta \tau$=0.2 ns). For the 
four $Q(t)$ signals in Fig. \ref{fig:12}, the estimation of time shifts are compatible with $\tau(k)$=$k\delta t$ ($k=1,...,4$) and for the 
signal $Q_2(t)$ $\tau$ is compatible with $\tau_2$=$\delta t/15$ measured more precisely elsewhere, all within the same $\sigma_t$ precision.
 
\subsection{the intensity spectrum in Zukunftoptik}
\label{subsec:OPSEC3_5} 

At this point we have drawn conclusions on the concept of intensity spectrum and seen that it is totally distinct from that of wavefunction 
spectrum. We made our case with the emission by a given point source at two slightly different angles of a common light signal varying at 
gigahertz frequencies, then received by two independent detectors and analysed with two digital processing systems. The overall system has to 
count photons simultaneously at frequencies orders of magnitude above what is possible today. However we proved the concept either by 
numerical simulation or by emulation (using real duplicated photon counts signals stored in a computer). That demonstration computes the 
variation of the correlation coefficient of both signals when a translation in time moves one signal respectively to the other, expressed by a 
mutual bilinear form called the intensity interference pattern. This pattern has a maximum frequency below one gigahertz, that is six orders 
of magnitude lower than the quantum mechanical interference pattern. This is perfectly summarized by \citet{ref:13} in two famous statements : 
<<Interference between two different photons never occurs>>, <<the superposition that occurs in quantum mechanics is of an essentially 
different nature from any occurring in the classical theory>>. More prosaically the analysis in this section emulates a classic gigahertz 
light intensity interferometry. In agreement with Dirac statements, this intensity interferometry is of different nature from the Michelson 
quantum mechanical interferometry. The common intensity spectrum (in fact a spectrogram) is conform to the Rayleigh, Plank, Einstein and von 
Laue's light spectrum defined in \citet{ref:1}, but it is ``of an essentially different nature'' from the quantum mechanical spectrum obtained 
by the elastic scattering of a single photon on a grating or through a prism (which is everyone else's optical spectrum).

\section{Electric current signals generated by light and RF intensity variations}
\label{sec:OPSEC4}

We announced in Section \ref{sec:OPSEC2} the possibility to measure, through a conversion into electric current, a classical 
``microscopic'' radiant intensity conform to the SI definitions but too small and varying too fast for the SI thermodynamic measurement 
method based on a microcalorimeter heat signal. In Section \ref{sec:OPSEC3} we used essentially the CCD \emph{photon-resolved} charge signals 
from \cite{ref:2}. Now in this section, we shall go into the details of the SI electric current measurement. We call it
\emph{voltage-resolved} electric current signals, like the 5 Ghz digitized signals produced by the setup of Fig. \ref{fig:1} which 
superposes light signals amplified by a PM\footnote{a few $10^5$ secondary electrons per photoelectron driven by $\approx100V$ to anode 
yielding each the current signal in Fig. \ref{fig:6}} and RF signals received by an antenna. Both signals are propagated together
along a coaxial line as an electric current. Then they are ``received'' by the digitizer which input impedance is ``matched'' with 
the line characteristic impedance and stored as an unique discrete digital time signal. The main practical difference between this sort of 
signals and the photon-counting ones is the possibility of measuring the time structure of the intensity variations at higher frequencies 
(several gigahertz), but the price to pay is the lack of a deep understanding of the electric current characteristics and that of its digital
sampling process.
Practically the periodic sampling of a periodic electric current source using a prior knowledge of its period allows the \emph{synchronous 
amplification} method, which averages the signal on a large number of periods -- 8192 for our usual DFT's -- (improving signal/noise by a factor 
90). This renders the periodic electric current signals in Section \ref{subsec:OPSEC4_3} visible in the frequency variable, even if they are 
\emph{hidden}\footnote{either hidden by the emission of another source or because its amplitude is lower than one adu} in the time variable. 
Lastly we compare in Section \ref{subsec:OPSEC4_4}, as we did in Section \ref{subsec:OPSEC3_3}, pairs of duplicated ``random noise'' electric 
current signals issued from a common radiant source, but using voltage-resolved instead of photon-resolved signals. This is the base of our original 
model of radio-astronomy interferometry : an intensity interferometry applied to two weak variable electric current time signals, usually hidden
deeply.

\subsection{digital electric current signals}
\label{subsec:OPSEC4_1}
The light and the RF photons are known by astronomy and radio-astronomy to propagate freely in vacuum, in straight line and at the universal light's speed $c$ on cosmic distances. 
When photons are converted into an electric current $I(t_k)$ and fed in a transmission line (a coaxial or a differential pair), this current propagates, guided by the line, at a fraction $(\epsilon \mu)^{-1/2}$ of $c$ and suffers resistive and guiding losses. 
This was approximately known already around 1860, from a study of the telegraph signal on long distances. 
The relevant model consisted then of a travelling pair of symmetric charge-density pulses creating a travelling transverse electric field between both conductors. 
\citet{Max1865} established a link between this travelling transverse electric field pulses and a light beam, which based his ETL concept. 
This corresponds to the PM electric current pulses guided by the transmission line of  Fig. \ref{fig:1}.
We have measured in Fig. \ref{fig:6}, relying on the superposition principle, the average DFT modulus of the billions of electric current pulses generated by all secondary electrons bunches with a $10^{-4}$ precision due to the high statistics. 
By inversion of this average DFT, we have recovered experimentally the average shape of these electric current pulses. 
This is exactly the Schottky's electric impulse response $F(t-t_k)$ in Eq. \ref{eq:eq1} (too small to be measured directly with a single primary photoelectron). 
In the same line of thought, in this section, we will analyse the correlation of whole electric current signals, instead of individual photons in a time series as done in Section \ref{sec:OPSEC3}. 
Their voltage time signal $R\times I(t)$ can be either processed analogically by elaborate analog circuits, or processed algebraically after being sampled/held/digitized. 

The first alternative corresponds to the traditional analog techniques of RF processing that have been developed for communication broadcasting, radar or radio-astronomy during a century (frequency filters, correlators, integrators, etc.). 

The second corresponds to fast digital methods introduced recently owing to the progress of nanoelectronics and of massive computing. 
They yield digitized superposable voltage-resolved electric current time signals with similar properties for light and for RF, but with different responses (the electron impulse $F(t-t_k)$ for light and the photon antenna response for microwave). 
The digital intensity waveform transmitted by the electric current before digitization is therefore a linear function of time resulting of the sum of the light intensity and of the RF intensity waveforms. 
This waveform of <<essentially  different nature>> from a photon wave-function, is 
stored, controlled, mitigated, processed, superposed and correlated to other waveforms generated by the same source system. 
It no more bounded by the power and the frequency limits of photon counting.
The typical exemples presented hereafter extend the RF and the light digital measurements of intensity above one gigahertz, only limited by the dynamic range of the analog processing electronics. 
The DFT linear algebraic operations apply to arbitrary\footnote{sine and cosine waves are matrix coefficients; taking them as physical components of intensity waveforms is a mistake} digital waveforms. 
They are fully reversible and have two main applications. 

First, in our example, using the frequency representation augments the signal/noise ratio at high frequency (cf. $\phi_0$ in  Fig. \ref{fig:1}).
Indeed, the noise intensity decreases by a factor thousand from 0.3 to 2500 Mhz, while in the time representation noise intensity is constant (cf. $\phi_0$ in   Fig. \ref{fig:3}). 

Second, as seen in  Fig. \ref{fig:9}, representing any time translation on arbitrary intensity waveforms is simple. 
This has served in Section \ref{subsec:OPSEC3_4} and will serve in Section \ref{subsec:OPSEC4_4} to correlate the signals received from the same source on two independent detectors. 

\subsection{periodic light intensity time signals}
\label{subsec:OPSEC4_2}

\begin{figure}
\resizebox{\hsize}{!}{\includegraphics{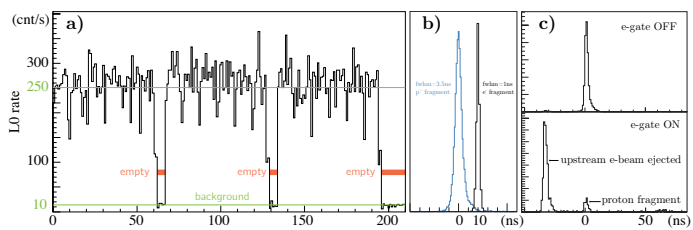}}
\caption{\textbf{a)} This histogram represents the L0 trigger rates ($\cong$250/s) for the sequence of bunches ($b=1-220$) of 96 ns constituting a full Hera ring circumference (including three groups of ``empty bunches'' in red counting only background triggers); \textbf{b)} proton's fragments (blue) hit Spacal back 10 ns before electron's fragments hit its front, with a longitudinal spread inherited from proton parents (3.5 ns vs 40 ps) ; \textbf{c)} top: fragment from p-gas collision hit ``hotspot'' counter; bottom: in-coming electron ejected from beam pipe before entering H1 or fragment from $e$-$p$ hit ``hotspot''}
\label{fig:14}
\end{figure}

Fast periodic light pulses are either driven by an electric pulse on a LED and guided by a fiber, or by a high energy particle beam radiation (the exemple presented here) or found in astronomy (e.g. the Crab pulsar). 
When they are photon-resolved, detectors limit useful fluxes (e.g. for PM $\phi$<$\phi_1$). 
The essential quality of the light intensity signals that have been presented previously was that they were photon-resolved. 
We have not found in astronomy such light signals varying in the gigahertz range. 
This rules out the possibility of distinguishing shot noise from genuine low level signal variations, which was the basis of absolute intensity calibration in \cite{ref:2}. 
We present in this section a periodic emission of correlated light pulses constituting many excellently reproducible and precise periodic time series. 
The primary sources of photons in our exemple are secondary high energy particles in the HERA electron-proton collider, with time of flight aligned with either proton or electron beams. 
These photons are detected by either the optimized Spacal PMs and their electronics \citep{ref:16}\footnote{efficient single photon detection, effective multiphoton time resolution of 0.1 ns from high B field mesh PM by CFD in ARC mode, 0.23 ns (RMS) time resolution for switching the sum of all 1500 Spacal channels energy when passing a virtual ``TOF'' gate separating proton secondaries from electron secondaries computed by adjusting 1500 digital delay lines}, 
or seen by other PMs near the vacuum beam pipe of Hera accelerators in the H1 experiment in a so-called ``L0 trigger'' test \citep{ref:15}. 
During this test the average L0 event rate of 28.4 kHz is unevenly distributed among the 220 ``Hera bunches'' (96 ns long each) as in seen in  Fig. \ref{fig:14}-a. 
Notably one sees three groups of ``empty bunches'', where the only L0 triggers are random ones (cosmic rays or radioactive decays). 
A full Hera interaction bunch corresponds to the crossing of 900 GeV protons trajectories through the 30 GeV electrons ones, both at light speed at the intersection of the two 1 km radii Hera rings at the center of the H1 detector. 
The position of head-electrons along their ring is defined by the Hera master clock edge at $\approx$0.3 mm (1ps) and the following ones are packed within 40 ps. 
These high precision results essentially of the quality of the control of the Hera $e$/$p$ beams mean trajectories using the electric signals from pickup electrodes along the rings.
We can reasonably compare this type of electric current (charged particles accompanied by the opposite polarity pickup current traveling at relativist speed on the beam pipe) to Maxwell's ``electro-magnetic wave'' model yielding symmetric currents in parallel or coaxial conductors (hence a transverse electric field), at a fraction of the light speed. 
Let us complete the description of Hera's periodic photon emission by the description of the photon hits in L0 PMs recorded by a multihit TDC chip synchronized with the Hera clock. 
This TDC chip, designed for the Stanford \emph{BaBar} experiment with a 16 ns bunch spacing, memorizes continuously the arrival time of 16 different signals with a 0.5 ns time resolution. 
It is continuously read, stored and analysed by our 29K fast processor. Each Hera period, which is conveniently 96 ns, an exact multiple of Babar period, is analysed independently from others with that 0.5 ns time resolution. 
Moreover the application of a 0.1 ns gate to the frontend signals of the incoming electrons permits their prior separation from protons circulating in the opposite direction. 
Fig. \ref{fig:14}-b shows the time distribution of proton and electron beams traversing spacal in the vacuum pipe. 
Flagging each Spacal hit by a fast coincidence (0.1 ns precision) with a time of flight ``$e$-gate'' and recording of the 0.23 ns rms mesh-PM signals with the TDC. 

\subsection{spectral analysis of hidden periodic RF intensity signals}
\label{subsec:OPSEC4_3}

We noticed in Section \ref{subsec:OPSEC2_2} that the many radio-frequency sources broadcasting in a few hundred meters from our test setup of  Fig. \ref{fig:1} appeared in that frequency representation as a forest of single frequency lines although they were hidden in the time representation of the same data by the zero-sum gaussian electronic noise variations of the ``dark'' pedestal $\phi_0$ in  Fig. \ref{fig:3}. 
The sources of these RF lines are the many RF photons detected by the virtual antenna and fed as an electric current in the coaxial line terminated by the amplifier-digitizer which yields a time series of $10^7$ successive electric current samples, a total limited by the size of the digitizer memory which corresponds to a 2 ms recording time. 
The same records as $\phi_0$ ``darks'' but with the PM exposed to six visible photon fluxes ranging from $\phi_1$=0.06 to $\phi_6$=2.5 p$e^-$/ns were analysed in Section \ref{subsec:OPSEC2_3} with the same method than used in the following but with much simpler results. 
These results, presented in  Fig. \ref{fig:6}, where focussed on stationary photon fluxes and ignored RF lines. 
A whole time series is subdivided in successive segments each transformed by DFT into two spectral representations : the power intensity spectrograms seen in Section \ref{subsec:OPSEC4_3_1} and the phase spectrograms in Section \ref{subsec:OPSEC4_3_2}. 

\subsubsection{the power spectrogram}
\label{subsec:OPSEC4_3_1}

\begin{figure*}
    \centering
    \includegraphics[width=17cm]{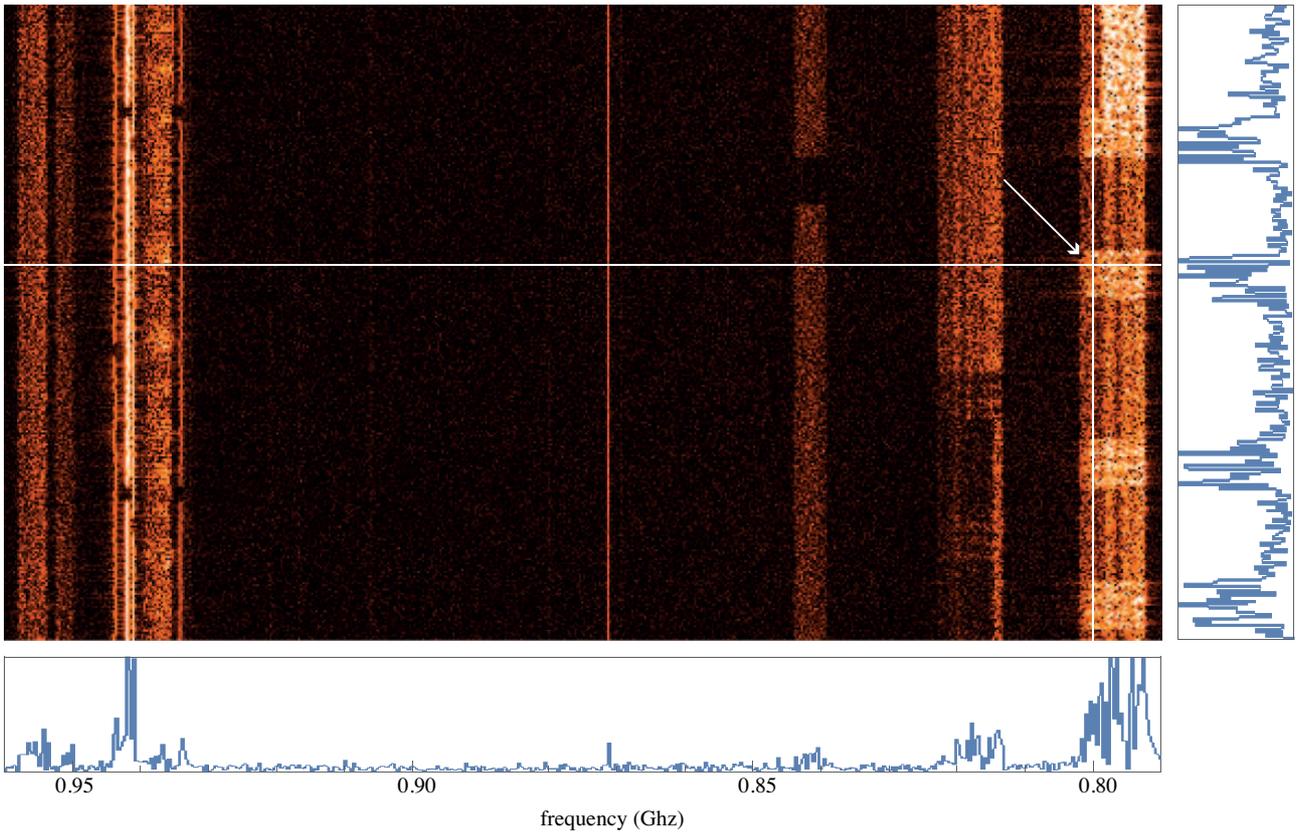}
    \caption{Spectrogram B (mid frequency): evolution of the intensity spectrum using data detected during 1 ms. Histograms are distributions in the line and column pointed by an arrow (white).}
    \label{fig:15}
\end{figure*}

\begin{figure}
\resizebox{\hsize}{!}{\includegraphics{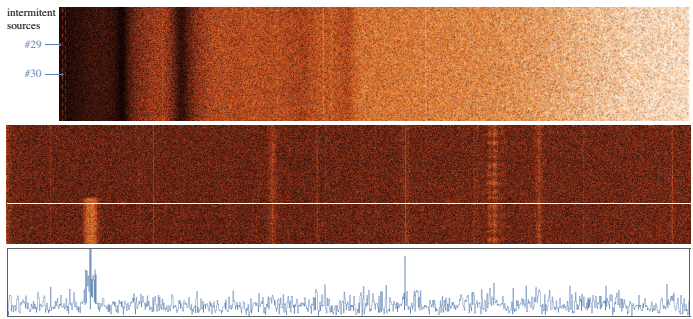}}
\caption{\textbf{top)} Spectrogram A ($\nu<0.45Ghz$ exposed to the photon flux $\phi_5$); \textbf{mid)} Spectrogram C ($\nu_N=2.5> \nu > 2Ghz$); \textbf{bottom)} power profile along the white line. Frequency axes are inverted by heterodyning.}
\label{fig:16}
\end{figure}

This power spectrogram uses the $\phi_5$ time series of $10^7$ intensity samples (8 bits) subdivided into 610 segments of $2\times8192$ samples (fine time sampling variable $\delta t$) each transformed by DFT into 610 pairs (power/phase) of spectrograms lines (coarse time sampling variable $\Delta t$) of 8192 columns (fine frequency sampling $\nu_N=1/2\delta t$). 
The dynamic range of the power samples is a factor $\approx$$10^5$. 
This power spectrogram is shown on a computer screen by the well-known imaging technique of astronomical CCD images (NASA's ``DS9''). 
Owing to its interactive features, DS9 gives access to a much larger quantity of information than what is shown in the 3\% extract in  Fig. \ref{fig:15} ($\phi_5$ spectrogram B). 
Nevertheless this extract, spanning only a 1 ms time and a 0.157 Ghz bandwidth around 0.87 Ghz in the overall intensity spectrum, shows the large quantity of information encoded inside the ``hidden'' microwave intensity data measured by our detector, memorized in our files and time-averaged in  Fig. \ref{fig:1}. 
It gives a good oversight of the complexity and the variety of modern digital radiosources and their high time-vs-frequency resolution seen in the profile histograms along the horizontal and vertical lines crossing at the point indicated by an arrow in Fig. \ref{fig:15}. 
These data are commonly considered as some ``pickup noise'' hidden by the electronic noise in the pedestal peak $\phi_0$ of  Fig. \ref{fig:3} and therefore not analysed. 
It is completed in  Fig. \ref{fig:16} by two other extracts A and C of the $\phi_5$ spectrogram, with a larger intensity frequency bandwidth, each representing 10\% of the overall power spectrogram. 
The horizontal axis of A represents frequencies decreasing from 0.45 Ghz to zero and the vertical axis covers 1 ms by steps of 3.3 \micro s. 
In this band the intensity spectrum is dominated by the stationary effect of the photon flux $\phi_5$ on the PM signal. 
More details about this stationary distribution and its shot noise (statistical fluctuation of anode electrons counts) are found in Section \ref{subsec:OPSEC3_2}. 
However we find a characteristic spectrum of high frequency microwave intensity variations, which pickup lines are visible, superposed to PM signal of the $\phi_5$ photon flux. 
In comparison, for the high frequency spectrogram C with no contribution of the PM signal and an electronic noise two order of magnitude lower than in A, the intensity detection has a much better intrinsic sensitivity. 
For these reasons we developed an automatic search of line or band emission sources effective on the whole zero to 2.5 Ghz domain by comparison with local background level at 3 RMS, more efficient than inspection of spectrograms by eye. 
A third of this microwaves sources, that are 34 in total, are confined in a single frequency channel numbered individually in a decreasing frequency order (most of them partly active during the eight 2 ms periods in our records). 
Another third of the sources are distributed in two adjacent channels and the rest presents a great variety as we see in  Fig. \ref{fig:15}. 
In order to conclude this RF power detection study, we should give our motivation to detect RF signals with a much lower power than what they are designed for (data transmission in the gigahertz range) : this is to get some practice with hidden electric current intensity signals considered in the next subsection to be the basic tool of the radio-astronomy interferometry.

\subsubsection{the phase spectrogram}
\label{subsec:OPSEC4_3_2}

\begin{figure}
\resizebox{\hsize}{!}{\includegraphics{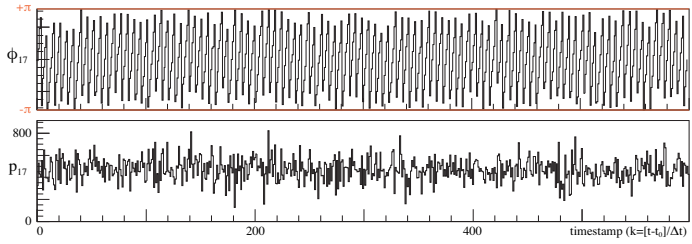}}
\caption{For the microwave source \#17 at 1.60 Ghz (one of the 34 single channel sources), phase $\phi_{17}(t)$ and power $p_{17}(t)$ as a function of time}
\label{fig:17}
\end{figure}

\begin{figure}
    \begin{center}    
\resizebox{6cm}{!}{\includegraphics{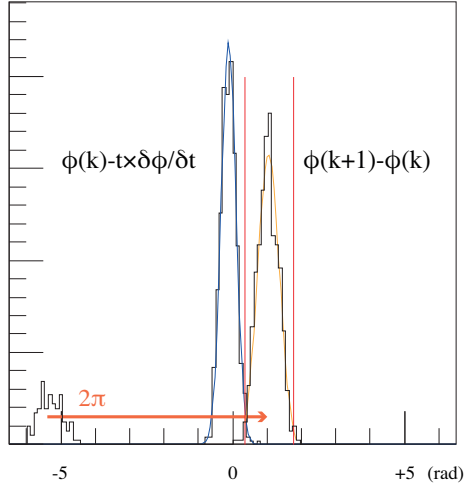}}
    \end{center}
\caption{Determination and fit of a continuous and linear phase vs time relation in the channel of source \#17 : the orange curve is obtained by 
fixing iteratively the phase ambiguity, aiming at a unique phase difference between consecutive phase samples; the blue curve results from a fit of 
phase samples vs time stamps points.}
\label{fig:18} 
\end{figure}

We restricted the study of the phase spectrograms to the 34 channels with a known single frequency power source. 
In these channels the phase is a continuous and linear function of time. 
For example we show in  Fig. \ref{fig:17} the phase $\phi_{17}(t)$ of source \#17 as a function of the coarse time variable $k\in \{1,...,610\}$ instead of the finer time variable $\delta t$ ($t=t_0+k\times \Delta t=t_0+k\times16384\times \delta t$) elapsed at the end of the recording of each spectrum. 
The source \#17, like the 34 others, is probably transmitting binary data continuously by emitting intensity peaks under the control of a very stable individual ``source clock''. 
This yields the detection of a phase variation linear with time when compared with the equally stable ``detector clock'' driving the sampling of the intensity signal by our waveform digitizer. 
We present a verification of this hypothesis by fitting a linear relation between phase and time instead of the seesaw seen in  Fig. \ref{fig:17}. 
This seesaw is due to the phase $\phi(t)$ reaching the $+\pi$ or $-\pi$ value, what leads to a $\pm 2\pi$ increment to restore continuity. 
The algorithm which cures this phase ambiguity created by Eq. \ref{eq:eq2} must 
\begin{enumerate}
    \item detect for which points on the raw $\phi(k)$ curve the phase needs to be incremented by a multiple of $\pm 2\pi$, then add the increments at these points; 
    \item fit a straight line on the incremented values of $\phi(k)$ and draw the two histograms controlling the result of the operation. 
\end{enumerate}
The process is presented on Fig.  \ref{fig:18}. The orange curve is the mean difference between successive values $\phi(k+1)-\phi(k)$ depending only on the constant slope of $\phi(t)$, that is $\delta f/\delta t\approx \langle\Delta \phi(t)/\Delta t\rangle=\langle\phi(t+\Delta t)-\phi(t)\rangle/\Delta t$.
The blue curve is the mean of the residuals $\langle\delta \phi_0(k)\rangle=\langle\phi(k)-(t-t_0)\times \delta \phi/\delta t \rangle$ obtained by the best fit of a straight line of slope $\delta \phi/\delta t$ on the rectified values of $\phi(k)$. It estimates the constant term of the linear relation of $\phi(t)$ vs $t$. 

The orange histogram in  Fig. \ref{fig:18}, obtained before the first iteration, shows that, in the case of the source \#17, a few values of the index $k$ are not correctly incremented. 
They appear in a peak with a phase shifted by $2\pi$  below the main peak. 
A first iteration increment the following phase samples leading to the unique blue peak. 
Extending this method to the 34 sources in our sample, we found 23 sources which have a complete linear phase fit after a number of iterations ranging from zero to fourteen. 
The other 11 sources are intermittent as seen by comparing the power signal with the phase signal in the same channel. 
Two typical exemples of intermittent sources (\#29 and \#30) are seen in spectrogram A (Fig. \ref{fig:16}), with on-and-off periods around 50 \micro s. 
During the off periods the phase is undefined. 
On the contrary when we consider the phase of the sources which like source \#17 have little power fluctuation, their phase $\delta \phi_0(k)$ behaves as a gaussian random variable. Their RMS $\sigma$ and the difference of the successive phases $\delta \phi_0(k+1)$ and $\delta \phi_0(k)$, measured independently as a slope, are a gaussian of RMS $\sigma \sqrt{2}$ (cf.  Fig. \ref{fig:18}). 
Otherwise if power fluctuations are larger, the fluctuations of $\delta \phi_0(k+1)$ and $\delta \phi_0(k$) can be correlated according to the frequency of intensity variations, that is the intensity spectrum. 
This behaviour is proved by studying the ratio of the width of the phase slope distribution to that of the single phase samples which varies from 1.4 ($\sqrt{2}$ ) down to 1.2 in our set of 23 steady sources. 
A part of this variation is understood by the correlation between the power and the phase in the same time sample. 
Another hidden part of this variation is likely originated in the time pattern of the emission which is averaged during the duration $\Delta t$ of the frequency analysis. 
A precise quantitative study of periodic signals requires a good control of the microwave periodic signal emission as well as transmission, signal sampling and digitization. 
This is a prerequisite for producing the high precision calibration tools that we need for a fundamental study of intensity spectral detectors.

\subsection{correlation between two electric current time-variations} \label{subsec:OPSEC4_4} 

We introduced the intensity interferometry as a correlation analysis of two time series representing  duplicated variable-intensity photon signals emitted continuously at two angles by the same point source.
Then, we took into account the translation due to the difference of optical path, in line with Rice's historical review of the intensity spectrum concept. 
Here we extend the correlation analysis to the electric current time series such as those used in Section \ref{subsec:OPSEC4_2} and Section \ref{subsec:OPSEC4_3}, also digitized in successive $\delta t$=0.2 ns intervals. 
With both opto-electric and radio-electric current series, we do not find any indication of either a time resolution limit or a non-linear response of the intensity detection at $\nu$=2.5 Ghz and below, but we lack for a real precise calibration such as in \cite{ref:2}. 
On these premises we understand the workings of radio-astronomic interferometry:  it checks the linear relation and the translation effect of duplicate intensity signals in the frequency variable, and in return, proves that these intensity signals are present in both duplicated electric currents. 
Technically we emulated this effect in Section \ref{subsec:OPSEC3_4} 
with two duplicated photon-resolved signals pushed to higher frequency by changing their time scale. 
The DFT of both time signals transformed this effect from the time variable (Fig. \ref{fig:9} and Fig. \ref{fig:10}) to the frequency variable (Fig. \ref{fig:11}). 
We reproduce this technique with the radio-electric currents from the previous paragraph by partitioning each current time signal into a sequence of segments each transformed by DFT into a pair of spectrograms (power $\times$ phase).
For example, with $10^7$ samples we form 610 consecutive segments each transformed into $2^{13}$ frequency channels (power $\times$ phase) pairs covering 3.28 \micro s out of a total of 2 ms time. 
This transformation was necessary to boost the frontend intensity initially below 0.4 adu by a signal/noise factor $90\approx\sqrt{2^{13}}$ at the 
expense of reducing the number of time samples by a factor $2^{14}$.
The two frequency spectra are overlaid and reach 2.5 Ghz frequencies. 
They are superposed to a stationary background with a constant intensity component thirty time larger than the variable part and affected by a stationary uncorrelated shot noise. 
This characterization fits to what is known about the emission by pointlike radio-sources superposed to the cosmic microwave background (CMB) and to galactic background. 
We also include the effect of the gaussian noise from digitization electronics which, in Radio-astronomy, may hide the electric current time signal.
The basic feature of the interferometric signal pattern in Section \ref{subsec:OPSEC3_4} is due to the translation in time between the intensity variations of two photon signal duplicates. 
This translation is caused either by a relative change of optical path length (rotation of the radiation source in Earth frame) or by a relative shift of the digitization clocks of both signal (phase switching method). 
This time translation is converted by DFT, in the frequency domain, into a relative phase shift varying linearly as a function of frequency. 
We did this operation on our digitized signals in a pure rigorous mathematical way with no loss of information, just by changing the sign of one half of samples -- the odd ones -- and performing the DFT on each of the 610 segments mentioned above. 
The \emph{heterodyning} operation is a Hadamard product with the Vandermonde DFT matrix.
This yields a new DFT matrix which produces the same frequency vector but in reverse order : the new low frequency spectrogram C  (Fig. \ref{fig:16}) has an almost null background.
The new high frequency spectrogram A has a high background (in our example the large intensity $\phi_5$).
The new Nyquist frequency component (the former average intensity) is totally out of scale.
In narrow bands (as those of Fig. \ref{fig:12}-b), this yields sine curve oscillations in the experimental interferometric data, as seen in Fig. \ref{fig:13}.
The fluctuation of these oscillations amplitude  results from the higher frequency component of the intensity signal and from the detector resolution. 
A basic difference between photon-resolved and current-resolved signals  remains. 
In common terms the first one is a DC yielding in Fig. \ref{fig:3} strictly negative photon amplitudes and the second one is an AC with no physical ``ground'' reference level. 
What we did in Fig. \ref{fig:15} and Fig. \ref{fig:16} is heterodyning.

\subsection{unification of light and of RF inside ZO} \label{subsec:OPSEC4_5} We have assembled some facts sufficient to 
integrate the signals conveyed by the photon fluxes of both light and RF in vacuum throughout spacetime in a common digital 
Zukunftoptik scheme presented for light in Section \ref{sec:OPSEC3}. This is part of the astronomy domain, but it was also 
tested in laboratory calibration benches. We bypassed the absence of a common emission frequency window for variable light 
and for RF sources in astronomy by shifting the time scale of our astronomical light records to transport them into the 
time scale of RF astronomy signals (0.2 ns/sample). The other important discrepancy to match was between our RF 
interferometry signals in \ref{subsec:OPSEC4_3} and \ref{subsec:OPSEC4_4} and the simplest ones that could reproduce the 
interferometry signals of the first generation of radio-interferometers in the nineteen-fifties. We know them  
from a detailed study of the first analog electronics experiments on point sources, before more 
refined experiments were done to yield two-dimensional images of these sources. In practice we digitized in Section 
\ref{subsec:OPSEC4_3} similar RF signals presenting the same two problems than RF point sources, being subliminal
(i.e. weaker than noise) and hidden (i.e. covered by an extended source stronger than the point-like target source). 
They justify a rigorous ZO model explaining the RF interferometry as a photon based intensity interferometry contrary to common
ETL explanations which are untenable. Similarly, a ZO model of light contrary to classic optics has been justified in Section 
\ref{sec:OPSEC3} at the unbeatable quantum precision limit, while RF cannot be studied at such a high precision.
However in light and RF, both components of ZO are present : a QM Born elastic scattering of photons on a concave mirror 
at $\approx$1 $eV/c $ for light and $\approx$0.1 $\mu eV/c$ for RF and a variable classical flux of particles averaged isochronously by the 
electric current in each detector.

\section{Mapping the universe with photons}
\label{sec:OPSEC5}

Mapping the universe with photons travelling throughout spacetime along straight rays, at any momentum, at a common speed $c$ in a dominantly 
dark and transparent vacuum\footnote{just containing other photons} is the program of astronomy and cosmology. We stated in the previous 
sections how this is done in the whole ZO band and how the number of photon rays passing through a source point and a detection point and its 
fluctuations are defined by QM and its statistics. This means that RF as well as light is carried by photons, not by some elusive 
<<electromagnetic waves>>. To complete this photon review, we shall resume in subsection \ref{subsec:OPSEC5.1} how 
easily astronomy has extended photon mapping to higher momenta and in Section \ref{subsec:OPSEC5.2} how one could try to extend photon mapping 
below ZO.

\subsection{high energy photons up to 1 PeV/c} \label{subsec:OPSEC5.1} This is the domains of X-ray and of Gamma-ray telescopes, each covering 
a bandpass larger than a $10^6$ factor, situated above the ZO band.  They use the photon-resolved electromagnetic calorimeters of particle 
physics with rates insufficient to envisage a correlation of two channels at the Ghz frequencies found in RF. This does not apply of course to 
the detection of a singular explosive astronomical nearby event (think of a nearby supernova). As we dont need to enter 
into the details of the great variety of high energy photon detectors used in this enormous domains, let us just note that the targets for 
X-ray telescopes are core atomic electrons in their various sub-bands (not the valence ones as in ZO) and for Gamma-ray telescopes the quasi 
free electrons or nuclei appearing when knocked at high energy. Therefore when high energy photons scatter elastically it is at a grazing 
incidence (what happens for gammas up to 10 Gev/c when targeting perfect crystals with sub-µrad wide photon beams). Consequently isochronous photon 
collection and imaging in a large field of view using macroscopic Born scattering on a concave reflector is not possible outside the ZO band.  
These high energy telescopes yield some information on a single photon momentum vector and the intensity of its source, but they lack the 
optical information on the photon ray and the quantum nature of its source. In particular they do not give access to the photon wavefunction 
which imposed itself in the ZO band.

\subsection{search of a low energy photon astronomy below ZO} \label{subsec:OPSEC5.2} Photon emitters and detectors exist in 
this domain. They are either quantized (Rydberg atoms, molecules, nuclear magnetic resonance) or not quantized (Compton and antennas) and 
they have found a great many applications such as transporting energy. 
However we found no examples of transmission of such relativist photon signals in astronomy at energy scales below RF. 
Among these lower levels of frequency and carrying energy in space, we find the <<current induction>> around 10 kHz or the 
<<extremely low frequency>> signal (ELF) around 50 Hz (6000 km wavelength) also emitted in space but trapped between the earth conductive layer 
and the ionosphere. 
Photon astronomy is based on the emission of photon rays propagating straight in vacuum at light speed $c$ and conserving the trace of their 
proper emission caracteristics in their wavefunction. 
However the above definition of an astronomic signal could fit to another recent example : the gravitational wave. It fits to the model of RF 
radio-astronomy interferometer given in Section \ref{subsec:OPSEC4_4} by locating the signal source by correlating the signals of three distant 
gravitational wave telescopes. This come naturally with a riddle : is the gravitational wave an extremely low energy photon 
signal? (or how would one prove that it is not?).

\subsection{electric fields and currents} \label{subsec:OPSEC5.3} When completing our search of a common <<photon>> model for astronomical signals 
travelling in vacuum, we left aside the concept of electromagnetic field. We propose to reintroduce field and current as a modernized version of 
Maxwell's Electromagnetic (EM) theory which has survived in all textbooks. This EM theory (essentially empirical) was born in the mid-nineteen 
century upon a common fit of the two separate empirical concepts of electric and of magnetic fields resulting of the early experiments with 
constants fields (Ampere, Faraday et al.), before \citet{Max1865} radical proposition to apply his EM theory to light (ETL). The 
practical application of the EM theory on stationary or slowly varying fields and currents measurements has been pushed to a high precision, 
integrated in the SI metrology and used ever since\footnote{while ETL remained in doubt}. It is reasonable to examine an eventual relation between 
such a stationary E field with the low energy photons met in subsection \ref{subsec:OPSEC5.2}, but without the rigorous proofs 
required in the rest of our analyses. To situate the framework of this question, it is no more a many light-year vacuum 
with an eventual few atoms thick metallic reflecting layer as in the rest of the article, but a solid state setup containing evacuated gaps 
thinner than the wavelength of the free photons that it contains. The borderline constant E-field setup consists of two opposite electric charges 
trapped in insulators creating some attractive or repulsive forces when maintained at a constant distance by a rigid frame. These forces are 
balanced by mechanical constraints transmitted by the frame elasticity. Similarly current loops yield B-field creating current-current 
inductive <<forces>> inside each loop or between them in a rigid setup balanced by the frame elasticity. One can imagine that, without 
a noticeable mechanical motion of condensed matter, inner mechanical stresses are caused by the recoils of the photon exchanges between 
charges. This quantification induces the classical quantification of the solid state with optical and acoustic phonons. Similarly with a more 
detailed analysis of setups like MRI imaging, polarized targets etc, one might prove the straight propagation of low energy photons rays 
without an artificial electromagnetic handwaving description of the setup.

\section{Conclusion}

\label{sec:OPSECCONCLUSION}

We introduced in this paper a new concept -- the dual component optics ``ZO''\footnote{with dual spectrum: wavefunction and intensity}-- and then, 
using the new technical means available for a few years, we searched the best ways to use each type of signal. 
First the single photon QM signal was already amply studied with CCD data in \cite{ref:2} under most aspects. The evolution of wavefunction 
during its propagation in vacuum was followed by moving a detector along the path of the beam. It remained to check in §\ref{subsec:OPSEC3_1}, that photon 
scattering on the edge of the telescope mirror integrates fully the diffraction effects inside the QM signal (while they are considered 
as independent in classical optics). 
   
Therefore the measurement of the second ZO component, the light intensity as a function of time, is the main task treated in this article using
firstly the photon count signal in Section \ref{sec:OPSEC3}, secondly the current detection sampling in Section \ref{sec:OPSEC4} and finally 
completing both methods by a spectral analysis. In the first case there are no light sources sufficiently bright to yield on astronomic or even 
on telescope scale distances a sufficient flux on the nanosecond scale and no detector to bear that flux. 

Using a ``state of the art'' $5 GHz$ current digitizer, reveals a forest of lines caused by ``picking-up'' subliminal RF photon signals emitted  
by a radio-transmitter in the neighborhood. That offered a downward extension of ZO wavefunction frequencies by a factor $10^7$. These photons
of 0.1 $\mu eV$ are not detectable, but the micro-currents from their periodic electric impulse response are piling up in the same frequency channel 
in the intensity spectrogram (cf. Fig.15 or Fig.16).  
These considerations lead us to propose an intensity correlation model of separate beams issued from a common RF point source hidden by an extended 
background as an explanation of the radio-astronomy interferometry. It fits to the simplest narrow band analog signal processing done at the birth 
of this field, in the nineteen-fifties (a product of two telescopes zero-suppressed signals from a single source moving in the field due to earth
rotation, i.e. a relative translation of both intensity signals). We take for granted that the complexification of signal and processing do not
change the nature of the RF sources detection and imaging.   

At last, we dealt shortly within Section \ref{sec:OPSEC5} with questions of a different nature, but that are needed for completion. 
We recalled that the photons with energy lower than RF exist, but not in an individually detectable or identifiable form. Moreover without any
knowledge of photon energy (whatever it means), we could envisage two other possible relativistic travelling candidates : dark energy and 
gravitational waves.

\begin{acknowledgements}
The author acknowledge XXX, YYY.
\end{acknowledgements}

\bibliographystyle{bibtex/aa} 
\bibliography{references}

\end{document}